%% file: TimeVary_NetFeat_Inference.tex
\documentclass[10pt,twoside]{IEEEtran}

\usepackage{amsfonts,amssymb,amsmath}
\usepackage{calc,bm,url,color,theorem,graphicx,cite,shortcuts_OPT,hyperref,cleveref}
\usepackage{psfrag,float,bbm}
\usepackage{algorithm}
\usepackage{algorithmic}
\usepackage{comment}

\usepackage[inline]{enumitem}
\def\inlinelabel{\sf{(\roman*)}}

\definecolor{orange}{RGB}{255,107,0}

\usepackage{color}
\usepackage[normalem]{ulem}

\newcommand{\MTSchange}[1]{{{\color{blue}#1}}}

\usepackage{mdframed}
\newmdtheoremenv{problem}{Problem}
\usepackage{microtype}

\newtheorem{Lemma}{Lemma}
\newtheorem{Prop}{Proposition}
\newtheorem{Theorem}{Theorem}

\theorembodyfont{\rmfamily}
\newtheorem{Exa}{Example}
\newtheorem{Remark}{Remark}
\newtheorem{Assumption}{Assumption}

\usepackage{pgfplots}
\usetikzlibrary{arrows,shapes,calc,tikzmark,backgrounds,matrix,decorations.markings}
\usepgfplotslibrary{fillbetween}
\usepgfplotslibrary{groupplots}
\pgfplotsset{compat=1.3}

\usepackage{relsize}
\tikzset{fontscale/.style = {font=\relsize{#1}}}

\definecolor{lavander}{cmyk}{0,0.48,0,0}
\definecolor{violet}{cmyk}{0.79,0.88,0,0}
\definecolor{burntorange}{cmyk}{0,0.52,1,0}

\definecolor{asuorange}{rgb}{1,0.699,0.0625}
\definecolor{asured}{rgb}{0.598,0,0.199}
\definecolor{asuborder}{rgb}{0.953,0.484,0}
\definecolor{asugrey}{rgb}{0.309,0.332,0.340}
\definecolor{asublue}{rgb}{0,0.555,0.836}
\definecolor{asugold}{rgb}{1,0.777,0.008}

\tikzstyle{stubborn}=[draw,circle, black!80, fill=black!40,
                       text=violet,minimum width=10pt]
\tikzstyle{superpeers}=[draw,circle, asublue!80!white, fill = asublue!50!white,
                       text=violet,minimum width=10pt]
\tikzstyle{susceptible}=[draw,circle, left color = red, color = red,
                       text=violet,minimum width=10pt]
\tikzstyle{perturb}=[draw,circle,burntorange, left color=blue,
                       text=violet,minimum width=10pt]
\tikzstyle{legend_general}=[rectangle, rounded corners, thin,
                           black, fill= white, draw, text=black,
                           minimum width=2.5cm, minimum height=0.8cm]
\tikzstyle{legend_graph}=[rectangle, rounded corners, thin,
                           black, fill= white, draw, text=black,
                           minimum width=1cm, minimum height=0.5cm]
\tikzstyle{legend_fw}=[rectangle, rounded corners, very thin,
                           black, fill= asuorange!8, draw, text=black]

\makeatletter
  \def\multilimits@{\bgroup
    \Let@
    \restore@math@cr
    \default@tag
    \baselineskip\fontdimen10 \scriptfont\tw@
    \advance\baselineskip\fontdimen12 \scriptfont\tw@
    \lineskip\thr@@\fontdimen8 \scriptfont\thr@@
    \lineskiplimit\lineskip
    \vbox\bgroup\ialign\bgroup\hfil$\m@th\scriptstyle{##}$\hfil\crcr}
  \def\Sb{_\multilimits@}
  \def\endSb{\crcr\egroup\egroup\egroup}
\makeatother

\makeatletter
\DeclareRobustCommand*\cal{\@fontswitch\relax\mathcal}
\makeatother

\begin{document}

\title{Exact Blind Community Detection from Signals on Multiple Graphs}
\author{
  T. Mitchell Roddenberry, Michael T. Schaub, Hoi-To Wai, and Santiago Segarra
  \thanks{T. M. Roddenberry and S. Segarra are with the Dept. of ECE, Rice University.
    M. T. Schaub is with the Dept. of ES, University of Oxford.
    H.-T. Wai is with the Dept. of SEEM, The Chinese University of Hong Kong.
    MTS received funding from the European Union’s Horizon 2020 research and innovation programme under the Marie Sklodowska-Curie grant agreement No 702410. 
    TMR partially received funding from the Ken Kennedy 2019/20 AMD Graduate Fellowship.
    The funders had no role in the design of this study; the results presented here reflect solely the authors’ views.
    Emails: mitch@rice.edu, michael.schaub@eng.ox.ax.uk, htwai@se.cuhk.edu.hk, segarra@rice.edu.
    Preliminary results appeared in a conference publication~\cite{Schaub2019}.}}

\date{\today}

\maketitle

\begin{abstract}
   Networks and data supported on graphs have become ubiquitous in the sciences and engineering.
   This paper studies the `blind' community detection problem, where we seek to infer the community structure of a graph model given the observation of independent graph signals on a set of nodes whose connections are unknown.
   We model each observation as filtered white noise, where the underlying network structure varies with every observation.
   These varying network structures are modeled as independent realizations of a latent planted partition model (PPM), justifying our assumption of a constant underlying community structure over all observations.
   Under certain conditions on the graph filter and PPM parameters, we propose algorithms for determining
   \begin{enumerate*}[label=\inlinelabel]
   \item the number of latent communities and
   \item the associated partitions of the PPM.
 \end{enumerate*}
   We then prove statistical guarantees in the asymptotic and non-asymptotic sampling cases.
   Numerical experiments on real and synthetic data demonstrate the efficacy of our algorithms.
   
\end{abstract}

\section{Introduction}
\label{sec:introduction}

The analysis of systems via graph-based representations has become a prevalent paradigm across science and engineering~\cite{Strogatz2001,Newman2010,Jackson2010}.
By representing a system as a graph, a plethora of system properties can be analyzed, including the importance of individual agents in the system \cite{pagerank1999}, the (possible) presence of a modular organization \cite{newman2006modularity}, and the prevalence of other connection motifs \cite{milo2002networkmotifs}.

However, while we may measure certain signals defined on the nodes or agents of the system, the edges coupling these agents are commonly unknown and need to be inferred from data.
This can be done either via an ad-hoc procedure such as thresholding a statistical association measure (\eg correlation or coherence) between the signals observed at the nodes \cite{bickel2009covariance}, or via more sophisticated statistical methods such as graphical LASSO and others~\cite{friedman2007lasso}.
While the former kind of approach has been successful in practice, it lacks theoretical guarantees as to its validity.
In contrast, a reliable exact inference of a graph requires a large number of independent samples, which are often not obtainable in practice.

Other issues faced in the inference of the exact network structure include fluctuations in the system structure itself: even though the large-scale features of the system are constant, the specific set of active edges between nodes may change with time or between realizations of the graph.
As concrete examples, consider observing the expression of opinions over time in a social network \cite{wai2016active,segarra2017consensus,shahrampour2013reconstruction,wu2018estimating,zhu2019network} or fMRI signals of different healthy patients in resting state~\cite{damoiseaux2006consistent,huang2018fmri}.
For the former case, individual active links might vary in each observation even when assuming a stable social fabric.
For the latter example, while individuals will have differing brain network structures at a fine scale, the large-scale network features will be similar.
As an additional example, consider observing daily stock returns in a market index.
There is a time-varying underlying network reflecting interactions between companies, which influences the price of individual stocks at a fine scale, even when the large-scale interactions between market sectors is stable~\cite{Hoffmann2018}.
We shall revisit this last example with numerical experiments in \Cref{sec:exp_stock}.

In the above scenarios, there is not a \emph{single} correct graph to be inferred and, therefore, any network inference method trying to find the \emph{correct} graph structure will fail.
However, certain features of the graphs may nevertheless be stable over each instance of the system and, thus, can be inferred from signals defined on these graphs.
Moreover, since features of common interest -- such as modular structure, centrality measures, and clustering coefficients -- are typically low-dimensional descriptions of the system (compared to the complete adjacency structure), we may recover these features directly from the observed signals with relatively few samples.

In this paper, \emph{we address the problem of inferring communities from the observation of data defined on the nodes of multiple (latent) graphs}.
Using the framework of graph signal processing, we model these data as graph signals induced by filters on the latent graphs \cite{segarra2017blind}.
In particular, we concentrate on the inference of 
\begin{enumerate*}[label=\inlinelabel]
\item the number of blocks and
\item the associated partitions of
\end{enumerate*}
a planted partition model.

\subsection{Related literature} 

The problem of network topology inference has been studied extensively in the literature from different perspectives including partial correlations~\cite{friedman_2008_sparse}, 
Gaussian graphical models~\cite{lake_2010_discovering, meinshausen_2006_high, egilmez_2017_graph}, 
structural equation models~\cite{bazerque_2013_sparse, baingana_2014_proximal}, 
Granger causality~\cite{sporns_2012_discovering}, and their nonlinear (kernelized) variants~\cite{shen_2017_kernel}. 
Recently, graph signal processing based methods for graph inference have emerged, which postulate that the observed data has been generated according to a network process defined on a latent graph~\cite{dong_laplacian_2016, kalofolias_smooth_2016, segarra_topo_2017, mateos_2019_connecting}.

While this work aims to recover structural properties of graphs from signals on the nodes, we do not recover the \emph{exact} structure of any graph.
Rather, we use graph signals to detect \emph{communities} of nodes, \ie sets of nodes with statistically similar connection profiles.
Community detection on graphs is a well-studied problem, typically seeking to partition the node set into blocks with a high density of edges within blocks and few edges between them \cite[Section \MakeUppercase{\romannumeral 2}-C]{Fortunato2016}.
Community detection methods include spectral clustering~\cite{von2007tutorial} that leverages approximately low-rank structures of connectivity-related matrices, statistical inference techniques that fit a generative model to the observed graph~\cite{ball2011,newman2007}, and optimization approaches that find communities that maximize modularity~\cite{newman2004}.
For an extensive review of community detection, we refer to~\cite{Fortunato2016}.

Our analysis focuses on the observation of signals supported by graphs drawn from a planted partition model (PPM), a popular graph model with ground-truth communities that has been extensively studied in the literature~\cite{holland1983,karrer2011,Abbe2018}.
In contrast to existing techniques for community detection, our problem formulation only observes signals on the nodes of a graph, rather than the set of edges.
This falls along the existing line of work on `blind' community detection, seeking to infer community structure in graphs with unknown edge sets.
Existing work in this direction considers the observation of node signals resulting from a diffusion process on a single graph~\cite{Schaub2018}, a low-rank excitation signal~\cite{Wai2018a}, or a transformation of a low-dimensional latent time series~\cite{Hoffmann2018}.
In contrast, our work considers the observation of signals over \emph{multiple} graphs, all drawn from the same latent PPM and each used to drive a network process yielding the corresponding observed signals.
We also note recent work on the blind inference of the eigenvector centrality for nodes in a graph, where~\cite{roddenberry2019ranking} considers the problem of ranking nodes according to their centrality in this regime, and~\cite{he2019centrality} considers the estimation of eigenvector centrality with colored excitations.

\subsection{Contributions and outline}
Our main contributions are as follows.
First, we provide an algorithm to detect communities from graph signals supported on a sequence of graphs generated by a PPM.
Second, we develop an algorithm to infer the number of groups in the underlying graph model from the observed signals.
Lastly, we derive both asymptotic as well as non-asymptotic statistical guarantees for both algorithms whenever the partition structure is induced from a planted partition model.
More specifically, we characterize the sampling requirements of both algorithms to achieve desired performance guarantees when finitely many samples are taken.

We first gather notation, background information, and preliminary results in Section~\ref{sec:sys}.
In Section~\ref{sec:problem_stat_and_algo}, we formally state the considered problems (Problems~\ref{P:main_problem2} and \ref{P:main_problem}) as well as the proposed algorithmic solutions (Algorithms~\ref{alg:model_select}~and~\ref{alg:timevary}), together with an illustrative example to highlight their effectiveness.
Our main theoretical results (Theorems~\ref{thm:1} and \ref{thm:2}) provide statistical guarantees for our algorithms, and are discussed in Section~\ref{sec:main_theory}.
Sections~\ref{sec:asymp} and~\ref{sec:nonasymp} contain the associated proofs of these results.
We complement our theoretical investigations with numerical experiments in Section~\ref{sec:numerical_exp}, before concluding with a short discussion highlighting potential avenues for future work in Section~\ref{sec:discussion}.

\section{Preliminaries: Graph Signal Processing and Random Graph Models}~\label{sec:sys}

\noindent \textbf{General notation.}
The entries of matrix $\bm{X}$ and (column) vector $\bm{x}$ are denoted by $X_{ij}$ and $x_i$, respectively.
For clarity, the alternative notation $[\bm X]_{ij}$ and $[\bm x]_i$ will occasionally be used for indexed matrices and vectors, respectively. 
The notation $^\top$, and $\mathbb{E}[\cdot]$ denote transpose and expected value, respectively.
$\bm{1}_k\in{\mathbb R}^k$ and $\bm I_k\in{\mathbb R}^{k\times k}$ refer to the all-ones vector and the identity matrix, and $\bm e_i$ denotes the $i$th standard basis vector.
For a given vector $\bm x$, $\diag(\bm x)$ is a diagonal matrix whose $i$th diagonal entry is~$x_i$.
The norm $\|\cdot\|_2$ indicates the $\ell_2\text{-norm}$ when the argument is a vector, and the induced $\ell_2$ operator norm when the argument is a matrix.
The notation $o(\cdot)$, $\mathcal{O}(\cdot)$, $\Omega(\cdot)$, and $\Theta(\cdot)$ take the established function approximation meaning~\cite[Chapter 3]{cormen2009algorithms}, with $\tilde{\Omega}(\cdot)$ denoting an approximation that ignores logarithmic terms.

\vspace{1mm}
\noindent \textbf{Graphs and graph shift operators.}
An~undirected graph $\mathcal G$ consists of a set $\mathcal N$ of $n:=|\mathcal N|$ nodes, and a set $\mathcal E$ of edges, corresponding to unordered pairs of elements in $\mathcal N$.
By identifying the node set $\mathcal N$ with the natural numbers $1,\ldots,n$, such a graph can be compactly encoded by a symmetric adjacency matrix $\bm A$, with entries $A_{ij}=A_{ji} = 1$ for all $(i,j)\in\mathcal E$, and $A_{ij} = 0$ otherwise.
Given a graph with adjacency matrix $\bm A$, the (combinatorial) graph Laplacian is defined as $\bm L := \bm D - \bm A$, where $\bm D = \text{diag}(\bm A\bm 1)$ is the diagonal matrix containing the degrees of each node.

The Laplacian and the adjacency matrix are two instances of a \emph{graph shift operator}~\cite{EmergingFieldGSP}.
A graph shift operator $\bm S \in \mathbb{R}^{n\times n}$ is any matrix whose sparsity pattern coincides with (or is sparser than) that of the graph Laplacian~\cite{EmergingFieldGSP}.
More precisely, $S_{ij} \geq 0$ only if $(i,j)\in \mathcal E$ or $i=j$, and $S_{ij}=0$ otherwise.
In this paper, we consider graph shift operators given either by the adjacency $\bm S = \bm A$ or the Laplacian $\bm S = \bm L$ matrices.
We denote the spectral decomposition of the graph shift operator by $\bm S = \bm V \bm \Lambda \bm V^\top$.
Here, $\bm V$ collects the eigenvectors of $\bm S$ as columns and $\mathbf{\Lambda} = \diag(\boldsymbol{\lambda})$ collects the eigenvalues $\boldsymbol{\lambda}=[\lambda_1,\cdots,\lambda_N]^{\top}$. 

\vspace{1mm}
\noindent \textbf{Graph signals and graph filters.}
We consider (filtered) signals defined on graphs as described next.
A \emph{graph signal} is a vector $\bm y \in \mathbb{R}^n$ that associates a scalar-valued observable to each node in the graph.
A \emph{graph filter} $\bm{\mathcal{H}}$ of order $T$ is a linear map between graph signals that can be expressed as a matrix polynomial in $\bm S$ of degree $T$:
\begin{equation}\label{E:filter_def}
    \bm{\mathcal{H}}(\bm S) = \sum_{l=0}^T h_l \bm S^l.
\end{equation}
For each graph filter, we define the (scalar) generating polynomial $h(\lambda) =  \sum_{l=0}^T h_l \lambda^l$. 

In this work, we are concerned with filtered graph signals of the form
\begin{equation} \label{E:g_sig}
    \bm y = \bm{\mathcal{H}}(\bm S) \bm w,
\end{equation}
where $\bm w$ is an excitation signal. 
By choosing appropriate filter coefficients, the above signal model can account for a range of  signal transformations and dynamics.
This includes consensus dynamics~\cite{Olfati-Saber2007}, random walks and diffusion~\cite{Masuda2017}, as well as more complicated dynamics mediated via interactions commensurate with the graph topology described by the Laplacian \cite{segarra_optimal_2017}.\vspace{.2cm}

\vspace{1mm}
\noindent \textbf{Stochastic block model.}
The SBM is a latent variable model that defines a probability measure over the set of unweighted networks of size $n$.
In an SBM, the network is assumed to be divided into $k$ groups of nodes.
Each node $i$ in the network is endowed with one latent group label $g_i \in \{1,\ldots,k\}$.
That is, if node $i$ is a member of group $j$, then $g_i=j$.
Conditioned on these latent group labels, each link $A_{ij}$ of the adjacency matrix $\bm A \in \{0,1\}^{n\times n}$ is an independent (up to symmetry of the matrix) Bernoulli random variable that takes value $1$ with probability $\Omega_{g_i,g_j}$ and value $0$ otherwise,
\begin{equation}
    A_{ij} | g_i, g_j \sim \text{Ber}(\Omega_{g_i,g_j}).
\end{equation}
To compactly describe the model, we collect all the link probabilities between the different groups in the symmetric affinity matrix $\bm \Omega \in [0,1]^{k \times k}$.
Furthermore, we define the partition indicator matrix $\bm G \in \{0,1\}^{n\times k}$ with entries $G_{ij} = 1$ if node $i$ belongs to group $j$ and $G_{ij}=0$ otherwise.
Based on these definitions, we can write the expected adjacency matrix under the SBM as
\begin{equation}\label{E:expected_G}
    \mathbb{E}[\bm A | \bm G] = \bm G \bm \Omega \bm G^\top.
\end{equation}

The planted partition model (PPM) is a particular case of the SBM, governed by two parameters $a,b$ for the link probabilities and having equally-sized groups.
In the PPM, the affinity matrix for a graph with $k$ groups and $n$ nodes can be written as
\begin{equation}\label{eq:ppmdef}
    \bm \Omega  = \frac{a}{n}\bm I_k  + \frac{b}{n}(\bm 1_k^{} \bm 1_k^\top - \bm I_k).
\end{equation}
Thus, the probability of an edge between any nodes within the same community is governed by the parameter $a$, whereas the probability of a link between two nodes of different communities is determined by $b$.

\section{Problem Statement and Algorithms}\label{sec:problem_stat_and_algo}
In this section, we formally introduce the two `blind' community detection problems to be studied.
We then specify the proposed algorithms to solve these problems, and provide an intuitive justification and an illustrative example.

\subsection{System model and problem statement}
As mentioned in Section~\ref{sec:introduction}, we are concerned with the inference of a statistical network model based on the observation of a set of \emph{graph signals}, \ie a scenario where the edges of the graph are unobserved. 
Consider a set of $m$ graph signals obtained as the outputs of graph filters.
For the $\ell$th instance, we observe $\bm y^\tl \in \RR^n$ given by [\lcf~\eqref{E:filter_def} and~\eqref{E:g_sig}]
\begin{equation}\label{eq:signal_model}
\bm y^{(\ell)} = \bm{\mathcal{H}}(\bm S^{(\ell)}) \bm w^{(\ell)}, \quad \ell = 1,\ldots,m.
\end{equation}
For every $\ell$, the graph shift operator $\bm S^\tl$ corresponds to an \emph{independently drawn PPM network with a constant parameter matrix $\bm \Omega$}.
The excitation signals $\bm w^\tl$ are assumed to be independent of the graph topology ${\bm S}^\tl$ as well as i.i.d. with zero mean and $\EE[ {\bm w}^\tl ( {\bm w}^\tl )^\top ] = {\bm I}$.

We introduce two specific `blind' problems that pertain to inferring properties of the unknown PPM \emph{without} observing the edges in the graphs drawn from this model.
The first goal is to select the \emph{model order}, specified as follows:
\begin{problem}[Model order selection]\label{P:main_problem2}
  Given a set of graph signals $\{{\bm y}^\tl\}_{\ell=1}^m$ following~\eqref{eq:signal_model}, infer the number of groups $k$ of the latent PPM generating $\bm{S}^\tl$.
\end{problem}
We estimate the number of blocks (or communities) in the graph in \Cref{P:main_problem2}.
This gives an initial coarse estimate of the structure of the network model. 

The estimated number of groups allows us to solve the following \emph{partition recovery problem}:
\begin{problem}[Partition recovery]\label{P:main_problem}
  Given a set of graph signals $\{{\bm y}^\tl\}_{\ell=1}^m$ following~\eqref{eq:signal_model} and a number of groups $k$, infer the community structure of the latent PPM generating~$\bm S^{(\ell)}$.
\end{problem}

There are several challenges related to solving Problems~\ref{P:main_problem2} and~ \ref{P:main_problem}. First, the graph topology is observed indirectly through the graph signals \eqref{eq:signal_model}.
Second, the graph topology is time-varying, as each ${\bm S}^\tl$ is drawn independently from the PPM.

\begin{Remark}
    We make the assumption of a white noise input with $\EE[ {\bm w}^\tl ( {\bm w}^\tl )^\top ] = {\bm I}_n$ in~\eqref{eq:signal_model} for simplicity, but successful recovery is possible even when this does not hold.
    This is illustrated in Section~\ref{sec:exp_colored}, where the system is excited by colored noise that is either not identical at every node, or has a rank-deficient covariance matrix.
\end{Remark}

\subsection{Proposed blind identification algorithms}

We summarize our proposed solutions to Problems~\ref{P:main_problem2} and~\ref{P:main_problem} in Algorithms~\ref{alg:model_select} and~\ref{alg:timevary}, respectively.
Both algorithms build upon the intuition that the spectral properties of the covariance matrix $\bm C_y = \EE[\bm{\mathcal{H}}(\bm S^{(\ell)}) \bm w^{(\ell)} \bm w^{(\ell)\top} \bm{\mathcal{H}}(\bm S^{(\ell)})^\top]$ of our observed signals will be shaped according to the block structure of the underlying PPM.
Note that the expectation is computed both with respect to the stochastic inputs $\bm w^{(\ell)}$ as well as the random graph $\bm S^{(\ell)}$.
To gain intuition, we first sketch the overarching algorithmic ideas and provide an illustrative example.
We postpone a more rigorous discussion to Sections~\ref{sec:asymp} and \ref{sec:nonasymp}.

From \eqref{eq:signal_model} it follows that for each instance $\ell$, the observed graph signal can be written as
\begin{equation} \label{eq:g_sig_decompose}
    {\bm y}^\tl = \left[ h_0\bm I + h_1 {\bm S}^\tl + \cdots + h_T ({\bm S}^\tl)^T \right] {\bm w}^\tl.
\end{equation}
Observe that the expectation of $[\bm S^{\tl}]^q$ will be low rank for every positive $q$, with a block structure inherited from the underlying PPM.
Since $\bm w^\tl$ is a white noise input, it can be shown (see~\Cref{sec:asymp}) that the covariance matrix $\bm C_y$ is given by a low-rank matrix plus a multiple of the identity matrix, thus inheriting the block structure from the PPM (see~\eqref{E:expected_G}).
As we do not have access to the true covariance matrix, both Algorithms~\ref{alg:model_select} and~\ref{alg:timevary} employ a sample covariance estimator instead of $\bm C_y$ as follows.

\paragraph{Model order selection} 

\algsetup{indent=1em}
\begin{algorithm}[tb]
	\caption{Model order selection algorithm}\label{alg:model_select}
	\begin{algorithmic}[1]
		\STATE {\textbf{INPUT}}: graph signals $\{\bm y^\tl\}_{\ell=1}^m$.
        \STATE Compute the vector of eigenvalues $\hat{\bm \lambda}$ of the sample covariance matrix  in descending order $\widehat{\lambda}_1 \ge \widehat{\lambda}_2 \ge \ldots \ge 0$:
		\begin{equation} \label{eq:cov}
            \textstyle \widehat{\bm C}_y^m \eqdef \frac{1}{m} \sum_{\ell=1}^m ({\bm y}^\tl ) ( {\bm y}^\tl )^\top. \vspace{-.4cm}
		\end{equation}
        \STATE Estimate $k^\star$ by finding the minimum description length:
        \begin{equation}
                k^\star = \argmin_{p\in\{1,\ldots,n\}} {\sf MDL}(p,\widehat{\bm \lambda}),
        \end{equation}
        with 
        \begin{align}\label{eq:mdl}
            \begin{split}
                {\sf MDL}(p,\widehat{\bm \lambda}) &:=  (p-n) \log \left( \frac{\prod_{j = p+1}^n \widehat{\lambda}_j^{\frac{1}{n-p}}}{\frac{1}{n-p} \sum_{j=p+1}^n \widehat{\lambda}_j} \right)\\ 
                &+ \hspace{0.2cm}  \frac{p}{2} ( 2n-p ) \frac{\log m}{m}.
            \end{split}
            \end{align}
		\STATE {\textbf{OUTPUT}}: estimated model order $k^\star$.
	\end{algorithmic} 
\end{algorithm}

Since the empirical sample covariance matrix $\widehat{\bm C}_y^m$ is only a (noisy) estimator of the true covariance, it will not be exactly low-rank plus identity, but will approximate this structure.
Intuitively, our aim is thus to estimate the model order $k$ such that the rank-$k$ approximation of $\widehat{\bm C}_y^m$ approximates the data well, while keeping $k$ as small as possible.
Algorithm~\ref{alg:model_select} achieves this by employing a minimum description length criterion $\text{MDL}(p)$ [\lcf~\eqref{eq:mdl}] to select an ``optimal'' number of groups $k$, see  \cite{wax1985detection}.

\paragraph{Partition recovery} 
Not only the eigenvalues of the sample covariance matrix, but also the eigenvectors carry valuable information to identify the underlying PPM.
It can be shown that the dominant $k$ eigenvectors of the (sample) covariance matrix are correlated with the partition structure of the PPM.
We can thus employ a procedure akin to classical spectral graph clustering~\cite{von2007tutorial,Rohe2011} to reveal the underlying partition structure via a $k$-means clustering of the rows of the matrix of dominant eigenvectors; see Algorithm~\ref{alg:timevary}.

\algsetup{indent=1em}
\begin{algorithm}[tb]
	\caption{Partition recovery algorithm.}\label{alg:timevary}
	\begin{algorithmic}[1]
		\STATE {\textbf{INPUT}}: graph signals $\{\bm y^\tl\}_{\ell=1}^m$ and nr. of blocks~$k$.
		\STATE \label{line:cov}Compute the eigenvalue decomposition $\widehat{\bm C}_y^m = \widehat{\bm V} \widehat{\bm{\Lambda}} \widehat{\bm V}^\top$ of the sample covariance matrix [cf.~\eqref{eq:cov}].
		\STATE Form $\widehat{\bm V}_k \in \RR^{n \times k}$ with the top-$k$ eigenvectors of $\widehat{\bm C}_y^m$.
		\STATE Apply $k$-means on the rows of $\widehat{\bm V}_k$. 
		\STATE {\textbf{OUTPUT}}: partition found by $k$-means.
	\end{algorithmic} 
\end{algorithm}

\subsection{Example: Inference with planted partition model}
\label{ssec:toy_example}
Let us illustrate the proposed algorithms with a simple example.
We generate synthetic data from a PPM with $n=100$ nodes, $k=2$ communities, $a = 4 \log n$, and $b = \gamma a$.
Note that for smaller $\gamma \in (0,1)$, the community structure in the randomly drawn graphs is more pronounced.
The input signal is uniform i.i.d.~with ${\bm w}^\tl \sim {\cal U}[-1,1]^n$ and the graph filter  is of the form ${\cal H}( {\bm L}^\tl ) = ({\bm I} - \beta {\bm L}^\tl )^5$, where
$\beta = 1 / (4 + 4\gamma) \log n$ and $\bm L^\tl$ is the Laplacian matrix of the $\ell$th sampled graph.

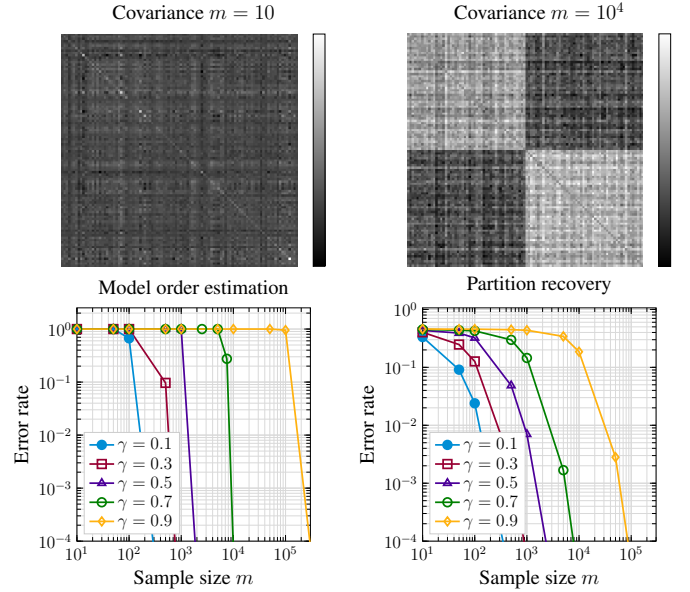
\begin{figure}[t]
\centering
\resizebox{\linewidth}{!}{\input{./example_experiments/ResLim_T1.tikz.tex}}
\vspace{-0.5cm}
\caption{Demonstration of algorithms for planted partition model. (Top) Snapshots of estimated covariance matrix $\widehat{\bm C}_y^m$, with nodes ordered for visibility: (left) $m=10$, (right) $m=10^4$.
  (Bottom) Error rates of solving the required tasks against sample size $m$ for synthetic graphs using Algorithms~\ref{alg:model_select} and \ref{alg:timevary}; see text: (left) model order estimation, (right) partition recovery.\vspace{-.2cm}}
\label{fig:reslim}
\end{figure}

Intuitively, for a large sample size, the sample covariance will be a good approximation of the true covariance matrix, thus, we should  obtain a satisfactory solution to our problems.
To illustrate this, in the top plots of Fig.~\ref{fig:reslim} we show a snapshot of the estimated covariance matrix $\widehat{\bm C}_y^m$ with sample sizes $m = 10$ and $m = 10^4$, with fixed $\gamma = 0.5$.
It is clear that with a large sample size such as $m = 10^4$, the estimated covariance matrix admits a clear partition into two blocks, which indeed correspond to the planted partition structure.
Note that the nodes are ordered in the plots for illustrative purposes, but in our observations we are not given such a convenient ordering.

The bottom plots in Fig.~\ref{fig:reslim} confirm the intuition that a larger sample size improves our estimates: here we plot the error rate of model order estimation and partition recovery (with known $k$) against the sample size $m$. 
For both problems, the error rates decay to zero as $m \rightarrow \infty$ regardless of the parameter $\gamma$. 
However, the error rate varies markedly over $\gamma$ when the number of samples is finite, highlighting the need for analysis of both asymptotic and non-asymptotic performance.

\section{Main Results}\label{sec:main_theory}
In this section, we describe our main theoretical results on the consistency and convergence rates of our algorithms to solve Problems~\ref{P:main_problem2} and~\ref{P:main_problem}.

For a generic graph filter ${\bm H} \eqdef \bm{\mathcal{H}}( {\bm S}^\tl ) \in \RR^{n \times n}$, the second order moments $\EE[ H_{il} H_{jl} ]$ can be characterized by the relative community membership of nodes $i,j,l$.
Using the expression $i \sim j$ to denote that both nodes $i$ and $j$ belong to the same group, whereas $i \not\sim j$ indicates the contrary, the possible values of the second order moments $\EE[ H_{il} H_{jl} ]$ can be specified by nine parameters:
\begin{align}\label{E:parameters_p_1_8}
&p_1 := \EE[H_{ii}^2], \, &&p_2 := \EE[H_{ij}^2]_{i \sim j}, \nonumber\\
&p_3 := \EE[H_{ij}^2]_{i \not\sim j}, \, &&p_4 := \EE[H_{ii} H_{ji}]_{i \sim j}, \nonumber\\
&p_5 := \EE[H_{ii} H_{ji}]_{i \not\sim j},  \, &&p_6 := \EE[H_{il} H_{jl}]_{i \sim j \sim l}, \nonumber\\
&p_7 := \EE[H_{il} H_{jl}]_{i \sim j \not\sim l}, \, &&p_8 := \EE[H_{il} H_{jl}]_{i \sim l \not\sim j}, \nonumber\\
&p_9 := \EE[H_{il} H_{jl}]_{i \not\sim l \not\sim j \not\sim i},
\end{align}
which depend on the graph filter and the PPM parameters $a, b$.
Furthermore define the following three parameters to simplify the presentation of our results.
\begin{equation}  \label{eq:c_def}
\begin{split}
& c_1\MTSchange{:}= 2 p_4 + \left(\frac{n}{k}-2 \right) p_6 + \frac{n}{k} (k-1) p_7, \\
& c_2 \MTSchange{:}= 2 p_5 + 2 \left(\frac{n}{k}-1\right) p_8 + \frac{n}{k} (k-2) p_9, \\
& c_3 \MTSchange{:}= p_1 + \left(\frac{n}{k}-1\right)p_2 + \frac{n}{k} (k-1) p_3. 
\end{split}
\end{equation}

\begin{Assumption}\label{ass:constants}
It holds for the parameters in \eqref{eq:c_def} that:
    \begin{equation} \label{eq:c_cond}
        c_3 > c_1 > c_2 \ge 0.
    \end{equation}
\end{Assumption}
In~\Cref{sec:exa} we show that Assumption~\ref{ass:constants} indeed holds for some common filter types.
Using the above notation, we now state the following results about the \emph{asymptotic} behavior of \Cref{alg:model_select,alg:timevary} when $m \rightarrow \infty$.

\begin{Theorem} \label{thm:1}
  Assume that there exists $r\geq 0$ such that $\|{\bm y}^\tl\|_2\leq\sqrt{r}$ almost surely and \Cref{ass:constants} holds.
  As the number of samples $m \rightarrow \infty$:
\begin{enumerate}
\item (\Cref{P:main_problem2}) Algorithm~\ref{alg:model_select} yields ${k}^\star = k$ w.h.p.
    \item (\Cref{P:main_problem}) \Cref{alg:timevary} recovers the true partition of the PPM w.h.p.
\end{enumerate}
\end{Theorem}
The proof can be found in Sec.~\ref{sec:asymp}.
We remark that the almost surely boundedness of $\| {\bm y}^\tl \|_2$ is guaranteed under mild conditions.
For example, it holds if
\begin{enumerate*}[label=\inlinelabel]
\item the spectral norm of the graph filter is bounded, \ie $\|\bm{\mathcal H} (\bm S^{(\ell)})\|_2 \leq \bar{h}$ for any $\ell$, and
\item the signal ${\bm w}^\tl$ is bounded.
\end{enumerate*}

In the non-asymptotic case when $m$ is finite, we have:
\begin{Theorem} \label{thm:2}
  Assume that there exists $r\geq 0$ such that $\|{\bm y}^\tl\|_2\leq\sqrt{r}$ almost surely and \Cref{ass:constants} holds.
\begin{enumerate}
\item (\Cref{P:main_problem2}) If $n, m$ are sufficiently large and
  \begin{equation} \label{eq:convbd1}
    m=\widetilde\Omega\left(\frac{1}{(c_3-c_1)^2}\right)
  \end{equation}
holds, then \Cref{alg:model_select} yields $k^\star\geq k$ w.h.p.
\item (\Cref{P:main_problem}) If
  \begin{equation} \label{eq:convbd2}
    m = \widetilde\Omega \left( \frac{1}{(c_1 - c_2)^2} \right)
  \end{equation}
holds, then \Cref{alg:timevary} {exactly solves} \Cref{P:main_problem} w.h.p.
\end{enumerate}
\end{Theorem}
The proof can be found in Sec.~\ref{sec:nonasymp}.
{Although the MDL criterion here is only guaranteed not to underestimate $k$ in Problem~\ref{P:main_problem2} when~\eqref{eq:convbd1} holds, we empirically observe in \Cref{ssec:exp_order_selection} that it tends to select $k^\star=k$ with sufficiently many samples, selecting $k^\star<k$ otherwise.}

Theorems \ref{thm:1} and \ref{thm:2} rely on the convergence of the covariance estimator \eqref{eq:cov}. To facilitate our discussions, let us borrow the following result from \cite[Corollary 5.52]{vershynin2010introduction}

\begin{Prop}  \label{prop:conc}
  Assume that there exists $r\geq 0$ such that $\|{\bm y}^\tl\|_2\leq\sqrt{r}$ almost surely.
  With probability at least $1-\delta$:
	\begin{equation} \label{eq:result1}
	\big\| \widehat{\bm C}_y^m - {\bm C}_y \big\|_2 \leq C_0 \sqrt{ \log(1/\delta) } \!~ \sqrt{ \frac{r}{m} }\eqs,
	\end{equation}
	where the constant $C_0$ satisfies $C_0 = \Theta( \| {\bm C}_y \|_2 )$.
\end{Prop}
It shows that the sample covariance converges at a rate of ${\cal O}( \sqrt{r/m} )$ for any fixed $\delta$.
These results can be generalized to the case of sub-gaussian ${\bm y}^\tl$, see \cite[Corollary 5.50]{vershynin2010introduction}.

Lastly, we remark that although we have focused on the special case of the PPM, the above results can be extended to the more general SBM model.

\subsection{Examples} \label{sec:exa}
Verifying \Cref{ass:constants} requires evaluating the second-order moments \eqref{E:parameters_p_1_8}, which is generally non-trivial. 
In Example~\ref{Ex:parameters_p}, we show that the simple graph filter given by the adjacency matrix of a PPM fulfills \Cref{ass:constants}.

\begin{Exa}\label{Ex:parameters_p}
	Consider the simple case where $\bm{\mathcal{H}}( {\bm S}^\tl ) = {\bm A}^\tl$, \ie the output of the underlying process at node $i$ corresponds to the sum of the initial values in the one-hop neighborhood of $i$. 
	For this case, we may explicitly compute the parameters in \eqref{E:parameters_p_1_8} to obtain $p_1 = p_2 = a$, $p_3 = b$, $p_4 = p_6 = a^2$, $p_5 = p_8 = ab$, and $p_7 = p_9 = b^2$, where we have allowed for self-loops in ${\bm A}^\tl$ to simplify the notation. 
	From~\eqref{eq:c_def} it then follows that the constants $c_i$ are given by $c_1 = \frac{n}{k} (a^2 + (k-1) b^2)$, $c_2 = \frac{n}{k}(2ab + (k-2) b^2)$, and $c_3 = \frac{n}{k}(a + (k-1)b)$.
        It follows immediately that Assumption~\ref{ass:constants} holds as long as $a \neq b$.
\end{Exa}

For the general case where the graph filter is \emph{any} polynomial of the adjacency matrix with positive coefficients, we can extend the above findings as illustrated in the next example.

\begin{Exa}\label{Ex:spectral_gap2}
Let $\bm{\mathcal{H}}( {\bm S}^\tl )$ be a polynomial of $\bm A^\tl$ with positive coefficients.
It can be shown that for a planted partition model with $a(n),b(n)\in o(n)$, $a(n) > b(n)$, and $k=2$ communities of size $n/2$ nodes each, the block structure of powers of $\bm A$ is maintained, \ie ${\mathbb E}([{\bm A}^t]_{ij})_{i\sim j}>{\mathbb E}([{\bm A}^t]_{ij})_{i\nsim j}\forall t>0$ for sufficiently large $n$, where we dropped the superscript $^\tl$ for clarity.
Thus, a positive coefficient polynomial $\bm{\mathcal H}({\bm A}^{(\ell)})$ maintains this block diagonally dominant structure in expectation.
Consequently, the covariance ${\bm C}_y$ is also block diagonally dominant, so we can leverage Proposition~\ref{P:limit_convergence} (to be stated in Section~\ref{sec:asymp}) to show that Assumption~\ref{ass:constants} holds.
\end{Exa}

\section{Proof of \Cref{thm:1}} \label{sec:asymp}
In this section, we prove our asymptotic consistency results stated in~\Cref{thm:1}.

We first characterize the spectral properties of the population covariance $\bm C_y$ (Proposition~\ref{P:limit_convergence} and Proposition~\ref{lem:eigvector}), and then discuss how the correct number of groups and the group memberships can be deduced from these properties.
We then use the fact that the sample covariance matrix $\widehat{\bm C}_y^m$ converges to ${\bm C}_y$ as ${m\rightarrow\infty}$, yielding the desired consistency guarantees.

\subsection{Spectral properties of population covariance}
We start by characterizing the covariance of the observed graph signals.
\begin{Prop}\label{P:limit_convergence}
  For a graph filter with parameters $c_1,c_2,c_3$ as defined in \eqref{eq:c_def}, it holds that 
	\begin{equation}\label{E:expected_value}
	{\bm C}_y = (c_3 - c_1) \bm I_n + \bm G \left( (c_1-c_2) \bm I_k + c_2 \mathbf{1}_k \mathbf{1}_k^\top \right)\bm G ^\top,
	\end{equation}
	where $\bm G \in \{0,1\}^{n\times k}$ is the partition indicator matrix as defined in \eqref{E:expected_G}.
\end{Prop}

\begin{IEEEproof}
    We show \eqref{E:expected_value} by explicitly computing the entries of ${\bm C}_y$ for a generic graph filter $\bm H \eqdef \bm{\mathcal{H}}( \bm S^\tl )$.
    The block structure of the PPM implies that we only have a finite set of cases to consider.

    First, we compute the diagonal entries of ${\bm C}_y$:
\begin{align}
\textstyle
[{\bm C}_y]_{ii} &= \EE[\bm h_i^\top \bm w \bm w^\top \bm h_i]= \EE\Big[ \Big(\sum_j H_{ij} w_j\Big)^2 \Big] \nonumber \\
&= \EE\Big[\sum_j H_{ij}^2 w_j^2 + \sum_{j,k} H_{ij} w_j H_{ik} w_i\Big] \nonumber \\
&= \sum_j \EE[H_{ij}^2] \EE[w_j^2] + \sum_{j,k} \EE[H_{ij} H_{ik}] \EE[w_j] \EE[w_i] \MTSchange{.} \nonumber
\end{align}
Using the fact that $\EE[w_j^2] = 1$ and $\EE[w_j]=0$, we have:
\begin{align}\label{E:proof_diag_element}
[{\bm C}_y]_{ii} &= \EE[H_{ii}^2] + \sum_{j : j \sim i} \EE[H_{ij}^2] + \sum_{j : j \nsim i} \EE[H_{ij}^2] \\
&= p_1 + \Big(\frac{n}{k}-1\Big) p_2 + \frac{n}{k} (k-1) p_3 = c_3. \nonumber
\end{align}
Second, we consider an off-diagonal entry in ${\bm C}_y$ \emph{within} a block of the PPM ($i \sim j$ but $i \neq j$).
\begin{align}
\textstyle
[{\bm C}_y]_{ij} &= \EE[\bm h_i^\top \bm w \bm w^\top \bm h_j]= \EE\Big[ \sum_{l,k} H_{il} w_l H_{jk} w_k\Big] \nonumber \\
& \overset{(a)}{=} \EE\Big[ \sum_l H_{il} H_{jl} w_l^2 \Big] \overset{(b)}{=} \sum_l \EE[H_{il} H_{jl}] \nonumber,
\end{align}
where (a) follows from $\EE[w_l w_k] = 0$ whenever $l \neq k$, and (b) follows from $\EE[w_l^2]=1$. 
We thus conclude that
\begin{align}\label{E:proof_off_diag_element_1}
[{\bm C}_y]_{ij} &= 2 \EE[H_{ii}H_{ji}] + \!\!\! \sum_{l : l \sim i, j \neq l \neq i} \!\!\! \EE[H_{il} H_{jl}] + \sum_{l : l \nsim i} \EE[H_{il} H_{jl}] \nonumber \\
&= 2 p_4 + \Big( \frac{n}{k} - 2 \Big) p_6 + \frac{n}{k} (k-1) p_7 = c_1.
\end{align}

Finally, considering $i$ and $j$ in different blocks of the PPM, we can show analogously that $[{\bm C}_y]_{ij} = c_2$. 
By combining this last result with \eqref{E:proof_diag_element} and \eqref{E:proof_off_diag_element_1}, expression \eqref{E:expected_value} follows.
\end{IEEEproof}

\begin{Remark}
  A similar result for the structure of the covariance matrix can be established if the underlying generating model of the graph is an SBM and not a PPM.
  However, the exact description of $\bm C_y$ will depend, in general, on all model parameters.
  For simplicity, we thus concentrate on the case of the PPM in this work.
\end{Remark} 
	
Proposition~\ref{P:limit_convergence} reveals the specific structure of the covariance matrix $\bm C_y$, which yields the following spectral properties.
\begin{Prop}\label{lem:eigvector}
Under Assumption~\ref{ass:constants}, the spectrum of ${\bm C}_y$ is characterized by:
\begin{equation} \label{eq:spectrum}
\begin{split}
& \lambda_{(1)} = c_3-c_1 + \frac{n}{k} (c_1 - c_2) + n c_2, \\
& \lambda_{(2)} = c_3-c_1 + \frac{n}{k} (c_1-c_2), \\
& \lambda_{(3)} = c_3-c_1,
\end{split}
\end{equation}
where $\lambda_{(i)}$ denotes the $i\text{th}$ largest eigenvalue.
The largest eigenvalue $\lambda_{(1)}$ has multiplicity one.
The eigenvalues $\lambda_{(2)}$ and $\lambda_{(3)}$ have multiplicity $k-1$ and $n-k$, respectively.

Moreover, the matrix of the top-$k$ eigenvectors $\bm V_k$ of $\bm C_y$ can be expressed as:
\begin{equation}
    \bm V_k = \bm G(\bm G^\top \bm G)^{-1/2} \bm U =: \widetilde{\bm G} \bm U,
\end{equation}
where $\widetilde{\bm G}$ is the normalized partition indicator matrix such that $\widetilde{\bm G}^\top\widetilde{\bm G} = \bm I$ and $\bm U$ is a unitary matrix.
\end{Prop}
\begin{IEEEproof}
From \eqref{E:expected_value}, we can see that the covariance ${\bm C}_y$  is composed of two terms, an identity matrix multiplied by a non-negative scalar $c_3-c_1$, and a rank-$k$ matrix 
$\bm F_k := \bm G \left( (c_1-c_2) \bm I_k + c_2 \mathbf{1}_k \mathbf{1}_k^\top \right)\bm G ^\top$.
Under Assumption~\ref{ass:constants}, $\bm F_k$ is positive semidefinite and therefore the top $k$ eigenvectors of $\bm C_y$ coincide with the $k$ eigenvectors of $\bm F_k$.

Let us define the diagonal matrix $\bm N_g := \bm G^\top \bm G$ and the matrix ${\bm \Phi:= \bm N_g^{1/2} \left( (c_1-c_2) \bm I_k + c_2 \mathbf{1}_k \mathbf{1}_k^\top \right) \bm N_g^{1/2}}$, such that $ \bm F_k = \widetilde{\bm G} \bm \Phi \widetilde{\bm G}^\top$.
Using the eigendecomposition $\bm \Phi = \bm U \bm \Lambda \bm U^\top$, it can be shown by direct computation that the matrix $\widetilde{\bm G}\bm U$ gathers the eigenvectors of $\bm F_k$:
\begin{equation*}
    \bm F_k \widetilde{\bm G}\bm U = \widetilde{\bm G} \bm \Phi \widetilde{\bm G}^\top\widetilde{\bm G}\bm U= \widetilde{\bm G} \bm \Phi \bm U= \widetilde{\bm G} \bm U \bm \Lambda.
\end{equation*}
Finally we note that since $\bm \Phi$ is symmetric, the matrix $\bm U$ is unitary.
It is then easy to verify that the eigenvectors $\widetilde{\bm G}\bm U$ are properly normalized.
\end{IEEEproof}

From the above result, it can be seen that the top-$k$ eigenvectors of $\bm C_y$ span $\mathrm{Im}(\bm G)$ and can therefore be used to recover the blocks of the underlying PPM.

\subsection{Establishing Theorem~\ref{thm:1}}
\begin{IEEEproof}[Proof of Theorem~\ref{thm:1} (\Cref{P:main_problem2})]
    From Propositions~\ref{prop:conc} and \ref{lem:eigvector}, we observe that the sample covariance matrix $\widehat{\bm C}_y^m$ converges to ${\bm C}_y$ as $m\to\infty$, and thus has three unique non-zero eigenvalues given by $\lambda_{(1)}, \lambda_{(2)}$, and $\lambda_{(3)}$ (in the limit).
    Moreover, under Assumption~\ref{ass:constants} the top $k$ eigenvalues of $\bm C_y$ are strictly larger than the lower $n-k$ eigenvalues.

    Accordingly, minimizing the MDL criterion yields $k^\star = k$.
    To show that the objective function ${\sf MDL}(p,{\bm \lambda})$ in \eqref{eq:mdl} has a minimum at $p=k$, we use~\eqref{eq:spectrum} and obtain for any $p \geq k$:
\begin{equation}\label{eq:mdlconsist}
    {\sf MDL}(p,{\bm \lambda}) = 
    \frac{1}{2} p (2n-p) \frac{\log m}{m},
\end{equation}
so ${\sf MDL}(k,{\bm \lambda}) < {\sf MDL}(k+1,{\bm \lambda}) < \cdots < {\sf MDL}(n,{\bm \lambda})$.

In addition, for any $2 \leq p \leq k-1$, we have
\begin{equation} \label{eq:mdlconsist2}
\begin{split} 
{\sf MDL}(p,{\bm \lambda}) & = 
\frac{1}{2} p (2n-p) \frac{\log m}{m} \\
& -(n-p) \log \left( \frac{ \lambda_{(2)}^{\frac{k-p}{n-p}} \lambda_{(3)}^{\frac{n-k}{n-p}}  }{ \frac{k-p}{n-p} \lambda_{(2)} + \frac{n-k}{n-p} \lambda_{(3)} } \right)  
\end{split}
\end{equation}
The fraction inside the logarithm can be expressed as
\begin{equation} \label{eq:taylor}
    \frac{ \big(  \lambda_{(2)} / \lambda_{(3)} \big)^{\frac{k-p}{n-p}}  }{ 1 + \frac{k-p}{n-p} \big( \frac{ \lambda_{(2)} }{ \lambda_{(3)} } - 1 \big)  } < 1,
\end{equation}
which holds as $\lambda_{(2)} > \lambda_{(3)}$ and $k < n$; see Appendix~\ref{app:small} for a detailed derivation.
This shows that the second term in \eqref{eq:mdlconsist2} must be a nonnegative number independent of $m$.

We thus conclude that the MDL criterion ${\sf MDL}(p,{\bm \lambda})$ attains its minimum at $p=k$ for the true eigenvalues $\bm \lambda$.
Finally, by Proposition~\ref{prop:conc} and Weyl's inequality, we have $\bm{\hat{\lambda}} \approx \bm \lambda$ as $m\rightarrow \infty$.
\end{IEEEproof}

\begin{IEEEproof}[Proof of Theorem~\ref{thm:1} (\Cref{P:main_problem})]
Denote the $i$th row vector of the eigenvector matrix $\bm V_k$ of $\bm C_y$  by ${\bm v}_i^{\rm row} \eqdef \bm V_k^\top \bm e_i$.
From Proposition~\ref{lem:eigvector}, we know that the matrix $\bm V_k$ has $k$ unique orthogonal row vectors (one for each group).
From~\Cref{prop:conc} we know that as $m \rightarrow \infty$ the empirical covariance matrix will converge to $\bm C_y$.
Hence, the vector $\bm v_i^\text{row}$ corresponding to node $i$ will correspond to one of those $k$ unique rows.
Clustering the vectors $\bm v_i^\text{row}$ into $k$ groups using $k$-means yields the desired partitioning.
\end{IEEEproof}

\section{Proof of \Cref{thm:2}}\label{sec:nonasymp}
The previous section shows the \emph{asymptotic} behavior of the proposed algorithm, when the covariance matrix is estimated perfectly.
In this section, we characterize the \emph{non-asymptotic} behavior of our algorithms in terms of the number of samples $m$ required to solve the considered problems with high probability.

\subsection{Proof of Theorem~\ref{thm:2} (\Cref{P:main_problem2})}
Our proof for \Cref{P:main_problem2} rests on two lemmas.
First, we bound the difference between the MDL criterion when applied to the empirical covariance ${\sf MDL}(p,\widehat{\bm \lambda})$ and the true covariance ${\sf MDL}(p,{\bm \lambda})$.
\begin{Lemma} \label{lem:mdlbd}
For any $p \in \{1,...,n\}$, it holds that
\begin{equation} \label{eq:mdlerr}
\big| {\sf MDL}(p,{\bm \lambda}) - {\sf MDL}(p,\widehat{\bm \lambda}) \big| \leq \frac{2 \sqrt{n}}{c_3-c_1} \| {\bm C}_y - \widehat{\bm C}_y^m \|_2 .
\end{equation}
\end{Lemma}
\begin{IEEEproof}
Recall the notation $\widehat{\lambda}_i = \lambda_i( \widehat{\bm C}_y^m)$ and $\lambda_i = \lambda_i ({\bm C}_y)$, and let $\delta \lambda_i = \widehat{\lambda}_i - \lambda_i $ be the estimation error of eigenvalue $i$. 
Observe that 
\begin{equation} \label{eq:mdl_diff_pf}
\begin{split}
& {\sf MDL}(p,\widehat{\bm \lambda}) - {\sf MDL}(p,{\bm \lambda}) \\
& = (n-p) \log \left( \frac{ \prod_{j = p+1}^n {\lambda}_j^{\frac{1}{n-p}} / {\prod_{j = p+1}^n \widehat{\lambda}_j^{\frac{1}{n-p}} } }{ \big(\sum_{j=p+1}^n {\lambda}_j \big) / \big(\sum_{j=p+1}^n \widehat{\lambda}_j \big) } \right) \\
& = (n-p) \log \left( \frac{ \prod_{j = p+1}^n  (1+ \frac{\delta \lambda_j}{\lambda_j} )^{-\frac{1}{n-p}} }{ \big(1 +  \frac{\sum_{j=p+1}^n \delta \lambda_j}{\sum_{j=p+1}^n \lambda_j} \big)^{-1} } \right).
\end{split}
\end{equation}
Taking absolute value on both sides of \eqref{eq:mdl_diff_pf} leads to 
\begin{equation}
\notag
\begin{split}
& |{\sf MDL}(p,\widehat{\bm \lambda}) - {\sf MDL}(p,{\bm \lambda})| \\
& \leq \sum_{j=p+1}^n \log \big( 1+ {\textstyle \frac{|\delta \lambda_j|}{\lambda_j}} \big) + (n-p) \log \Big( 1 + {\textstyle \frac{\sum_{j=p+1}^n |\delta \lambda_j|}{\sum_{j=p+1}^n \lambda_j}} \Big) \\
& \overset{(a)}{\leq} \sum_{j=p+1}^n \frac{|\delta \lambda_j|}{\lambda_j} + (n-p) \frac{\sum_{j=p+1}^n |\delta \lambda_j|}{\sum_{j=p+1}^n \lambda_j} \leq \frac{2}{\lambda_{(3)}}\sum_{j=p+1}^n|\delta \lambda_j|,
\end{split}
\end{equation}
where (a) is due to $\log(1+x)\leq x$ for any $x > -1$.

To simplify this expression further, we make use of the following generalization of Weyl's inequality:
\begin{equation} \label{eq:weyl}
\sum_{j \in {\cal S}} | \lambda_j( \widehat{\bm C}_y^m ) - \lambda_j( {\bm C}_y ) | \leq \sqrt{|{\cal S}|} \big\| \widehat{\bm C}_y^m - {\bm C}_y \big\|_2,
\end{equation}
which holds for any ${\cal S} \subseteq \{1,...,n\}$ and follows from \cite[Corollary 6.3.8]{horn2012matrix} and the equivalence of norms.

Employing \eqref{eq:weyl} over the set ${\cal S}=\{p+1,\ldots,n\}$, and plugging in the value of $\lambda_{(3)}$ as given in \Cref{lem:eigvector} yields
\begin{equation}
\notag
\begin{split}
& |{\sf MDL}(p,\widehat{\bm \lambda}) - {\sf MDL}(p,{\bm \lambda})| \leq \frac{2\sqrt{n-p}}{c_3-c_1}\|\widehat{\bm C}_y^m-{\bm C}_y\|_2,
\end{split}
\end{equation}
from which \eqref{eq:mdlerr} immediately follows.
\end{IEEEproof}

Second, we bound the difference between the MDL criterion when $p< k$ and $p=k$, both with respect to the true covariance.
\begin{Lemma}\label{lem:mdl_gap}
Let $\rho \eqdef \frac{1}{k}\frac{c_1-c_2}{c_3-c_1}$.
If $p < k$, then
\begin{equation} \notag 
\begin{split}
  {\sf MDL}(p,{\bm \lambda}) - {\sf MDL}(k,{\bm \lambda})
  \geq C_1 n \left(\log(1+\rho)-\frac{\log m}{m}\right).
\end{split}
\end{equation}
for some constant $C_1$ which depends on $n,k,\rho$.
\end{Lemma}

\begin{IEEEproof}
If $p < k$, we observe that the first term in the MDL criterion \eqref{eq:mdl} is monotonically decreasing in $p$, so it suffices to lower bound the difference by evaluating ${\sf MDL}(k-1,{\bm \lambda})-{\sf MDL}(k,{\bm \lambda})$.
This expression evaluates to
\begin{equation} \notag
\begin{split}
& {\sf MDL}(k-1,{\bm \lambda})-{\sf MDL}(k,{\bm \lambda}) = -\left(n-k+\frac{1}{2}\right)\frac{\log m}{m} \\
& \hspace{1.5cm}+ \log \left( \frac{(1 + \frac{1}{n-k+1}( \frac{\lambda_{(2)}}{\lambda_{(3)}} - 1 ))^{n-k+1}}{\lambda_{(2)} / \lambda_{(3)}} \right). \\
& =\left(n-k+1\right)\log\left(1+\frac{n}{n-k+1}\rho\right)-\log\left(1+n\rho\right) \\
& \hspace{1.5cm}-\left(n-k+\frac{1}{2}\right)\frac{\log m}{m}.
\end{split}
\end{equation}
Observing that $\log(1+n\rho)\in o(n\log(1+\rho))$, as well as $n\gg k$, the conclusion of the lemma follows.
\end{IEEEproof}

With these two results in place we can now conclude our proof of Theorem~\ref{thm:2} (\Cref{P:main_problem2}).
\begin{IEEEproof}[Concluding the proof]
\Cref{lem:mdl_gap} provides a lower bound on ${\sf MDL}(p,{\bm \lambda}) - {\sf MDL}(k,{\bm \lambda})$ when $p<k$.
If this lower bound is violated, then $k^\star\geq k$ must hold.

Combining \Cref{lem:mdlbd} with \Cref{prop:conc}, it suffices to show that $\frac{\sqrt{n}}{c_3-c_1} C \sqrt{r/m}$, for some constant $C$, violates the lower bound in~\Cref{lem:mdl_gap}.
The required condition can be written as
\begin{equation} \label{eq:thm2prob2lb1}
  \begin{split}
    & \frac{C \sqrt{r}}{c_3-c_1} \leq C_1\sqrt{n m} \left( \log( 1 + \rho ) - \frac{\log m}{m} \right) \\
    & \Longleftrightarrow \frac{C/C_1}{c_3-c_1}\sqrt{\frac{r}{n}} \leq \sqrt{m}\log(1+\rho) - \log{m}.
  \end{split}
\end{equation}
The desired sampling requirement~\eqref{eq:convbd1} follows.
\end{IEEEproof}

\subsection{Proof of Theorem \ref{thm:2} (\Cref{P:main_problem})}
To prove the second part of Theorem \ref{thm:2} related to \Cref{P:main_problem}, we need to bound the labeling error we obtain from applying $k$-means to the rows of the eigenvectors of the sample covariance matrix.
To this end, we proceed in three steps.

Let $\widetilde{\bm G}$ (resp.~$\widetilde{\bm G}_{\sf true}$) be the normalized indicator matrix induced by the candidate labeling $g$ (resp.~the true community labeling $g_{\sf true}$).
Consider the $k$-means objective function:
\begin{equation}\label{eq:costfct}
  F(g,\bm V_k) \eqdef \left \| [\bm I_n - \widetilde{\bm G}\widetilde{\bm G}^\top] \bm V_k \right\|^2_\text{F}.
  \end{equation}
Our first step in the analysis is to lower bound $F(g,\bm V_k)$ when ${\bm V}_k$ is selected as the top eigenvectors of the true covariance matrix $\bm C_y$.

\begin{Lemma} \label{lem:lowerbd}
 Assume that the true $k$ communities are of equal size with $\geq 2$ nodes.
 If there is no labeling error of the nodes, then  $F(g,\bm V_k) = 0$.
 Otherwise, we have that
\begin{equation}\label{eq:perturb_bd}
F(g,\bm V_k) \ge \ds {\frac{2}{n/k+1}}.
\end{equation}
\end{Lemma}

\begin{IEEEproof}
From \Cref{lem:eigvector}, we know that $\bm V_k = \widetilde{\bm G}_{\sf true}\bm U$.
Moreover, as $\widetilde{\bm G}_{\sf true}\widetilde{\bm G}_{\sf true}^\top = \widetilde{\bm G}_{\sf true}\bm\Theta\bm\Theta^\top \widetilde{\bm G}_{\sf true}^\top$ for any group permutation matrix $\bm \Theta$, it follows that $F(g,\bm V_k) = 0 $ if $g_i=\sigma(g_i^{\sf true})$ for some permutation map $\sigma : [1,\ldots,k] \rightarrow [1\ldots,k]$.

Next, we consider the case when $g$ mislabels at least one node.
We have the following chain of equivalence for $F$:
\begin{equation}
\begin{split}
& F(g,{\bm V}_k) = \| {\bm V}_k - \widetilde{\bm G} \widetilde{\bm G}^\top {\bm V}_k \|_{\rm F}^2 \\
& = \| {\bm V}_k \|_{\rm F}^2 - 2 \langle {\bm V}_k, \widetilde{\bm G} \widetilde{\bm G}^\top {\bm V}_k  \rangle + \| \widetilde{\bm G} \widetilde{\bm G}^\top {\bm V}_k \|_{\rm F}^2.	
\end{split}
\end{equation}
Notice that $\| \widetilde{\bm G} \widetilde{\bm G}^\top {\bm V}_k \|_{\rm F}^2 = \langle {\bm V}_k, \widetilde{\bm G} \widetilde{\bm G}^\top {\bm V}_k  \rangle$ 
since $\widetilde{\bm G}^\top \widetilde{\bm G} = {\bm I}$. This gives,
\begin{equation} \label{eq:cost_fct_almost_simple}
F(g,{\bm V}_k) = \| {\bm V}_k \|_{\rm F}^2 - \|\widetilde{\bm G}\widetilde{\bm G}^\top{\bm V}_k\|_{\rm F}^2.
\end{equation}
Recalling from Proposition~\ref{lem:eigvector} that ${\bm V}_k=\widetilde{\bm G}_{\sf true}{\bm U}$, this can be be further simplified to
\begin{equation}\label{eq:eq_cost_fct_simple}
F(g,{\bm V}_k) = k-\|\widetilde{\bm G}^\top\widetilde{\bm G}_{\sf true}\|_{\rm F}^2.
\end{equation}
Observe that
\begin{equation}\label{eq:cost_fct_entries}
  [\widetilde{\bm G}^\top\widetilde{\bm G}_{\sf true}]_{ij}=
  \frac{|\{x\in{\mathcal N}\colon g_x=i\text{ and }g_x^{\sf true}=j\}|}{\sqrt{s_is_j^{\sf true}}},
\end{equation}
where $s_i$ and $s_i^{\sf true}$ indicate the size of the $i$th partition under $g$ and $g^{\sf true}$, respectively.

Consider the case when $g$ mislabels exactly one node.
Without loss of generality, this scenario can be captured by mislabeling node $1$.
We have $g_1=1$, $g_1^{\sf true}=2$, and
\begin{equation}
  g_i = g_i^{\sf true},~~\text{if}~~i \neq 1.
\end{equation}
Using \eqref{eq:cost_fct_entries}, it can be shown by direct calculation that
\begin{equation}\label{eq:app-proof-lowerbd}
  F(g,{\bm V}_k)=\frac{s_1^{\sf true}+s_2^{\sf true}}{(s_1^{\sf true}+1)s_2^{\sf true}}.
\end{equation}
Under the assumption that the $k$ communities are of equal size, \ie $s_i^{\sf true}=n/k\ \forall i\in[1,\ldots,k]$, this directly yields the right-hand side of \eqref{eq:perturb_bd}.
We remark that minimizing~\eqref{eq:app-proof-lowerbd} over the possible choices of $s_1^{\sf true},s_2^{\sf true}$ will yield a lower bound for a PPM where the communities are \emph{not equally sized}.

We conclude the proof by noting that the cost function is increasing in the number of errors made by the labeling function, so \eqref{eq:perturb_bd} presents a lower bound for the cost when the candidate labeling is incorrect.
\end{IEEEproof}

{Before proceeding, we first note that when the sample bound~\eqref{eq:convbd2} is attained, the following holds as a direct result of Propositions~\ref{prop:conc}~and~\ref{lem:eigvector}:}
\begin{equation} \label{eq:s_gap}
\nu := \lambda_{(2)} - \lambda_{(3)} - \| {\bm C}_y - \widehat{\bm C}_y^m \|_2 > 0 \eqs.
\end{equation}

We then bound the suboptimality of the communities found by \Cref{alg:timevary} as follows.
\begin{Prop} \label{prop:exact}
  Under Assumption~\ref{ass:constants}, assume that \eqref{eq:s_gap} holds.
  If Algorithm~\ref{alg:timevary} returns an optimal $k$-means solution, then the following holds:
\begin{equation}\label{eq:d_bd}
F(g,{\bm V}_k) \leq \frac{4k}{\nu^2} \|{\bm C}_y - \widehat{\bm C}_y^m\|_2^2 \eqs,
\end{equation}
where $g$ indicates the communities returned by Algorithm~\ref{alg:timevary}.
\end{Prop}

\begin{IEEEproof}
  Given a candidate partition with labeling $g$, we define the normalized indicator matrices $\widetilde{\bm G},\widetilde{\bm G}_{\sf true}$ as in the proof of Lemma~\ref{lem:lowerbd}.

Let ${\bm E} \eqdef {\bm V}_k {\bm V}_k^\top - \widehat{\bm V}_k \widehat{\bm V}_k^\top $ be an error matrix. We observe the chain 
\begin{equation} \label{eq:l_chain}
\begin{split}
& F(g,{\bm V}_k) \\
& = \| ( {\bm I} - \widetilde{\bm G} \widetilde{\bm G}^\top ) {\bm V}_k \|_{\rm F}^2 \overset{(a)}{=} \| ( {\bm I} - \widetilde{\bm G} \widetilde{\bm G}^\top ) {\bm V}_k {\bm V}_k^\top \|_{\rm F}^2  \\
& \leq \left(\| ( {\bm I} - \widetilde{\bm G} \widetilde{\bm G}^\top ) \widehat{\bm V}_k\widehat{\bm V}_k^\top \|_{\rm F}  + \| ( {\bm I} - \widetilde{\bm G} \widetilde{\bm G}^\top ) {\bm E} \|_{\rm F}\right)^2 \\
& \overset{(b)}{\leq} \left(\| ( {\bm I} - \widetilde{\bm G}_{\sf true} \widetilde{\bm G}_{\sf true}^\top ) \widehat{\bm V}_k \widehat{\bm V}_k^\top \|_{\rm F} + \| {\bm E} \|_{\rm F}\right)^2 \\
& \overset{(c)}{\leq} 4 \| {\bm E} \|_{\rm F}^2,
\end{split}
\end{equation}
where (a) is due to ${\bm V}_k^\top {\bm V}_k = {\bm I}$,
(b) is due to $g$ yielding an
optimal $k$-means solution given $\widehat{\bm V}_k$, and
(c) used $F(g^{\sf true},{\bm V}_k) = 0$ together with the triangle inequality. 

Furthermore, we observe that the error $\bm E$ between the top-$k$ eigenvectors
of ${\bm C}_y$ and $\widehat{\bm C}_y^m$ is bounded by
\begin{equation} \label{eq:err-eigvec-bound}
\begin{split}
& \| {\bm E} \|_{\rm F} = \| {\bm V}_k {\bm V}_k^\top - \widehat{\bm V}_k \widehat{\bm V}_k^\top \|_{\rm F}  \overset{(a)}{\leq} \sqrt{k} \| {\bm V}_k {\bm V}_k^\top - \widehat{\bm V}_k \widehat{\bm V}_k^\top \|_2 \\
& = \sqrt{k} \| {\bm V}_k^\top \widehat{\bm V}_{n-k} \|_2  {\overset{(b)}{=} \sqrt{k} \left\| \sin \Theta\left( {\bm V}_k,\widehat{\bm V}_k \right) \right\|_2},
\end{split}
\end{equation}
where (a) is due to \cite[Lemma 7]{boutsidis2015spectral}, $\widehat{\bm V}_{n-k}$ denotes the matrix formed by the eigenvectors of $\widehat{\bm C}_y^m$ orthogonal to $\widehat{\bm V}_k$, {and (b) is due to a result noted in~\cite{daviskahan1970}.}
Define $\bm{\Delta} \eqdef {\bm C}_y - \widehat{\bm C}_y^m$. 
The $(k+1)$-largest eigenvalue of 
$\widehat{\bm C}_y^m$ \emph{does not} exceed $\lambda_k( {\bm C}_y )$:
\begin{equation}
\lambda_{k+1} ( \widehat{\bm C}_y^m ) \leq \lambda_{k+1} ( {\bm C}_y ) + \| \bm{\Delta} \|_2 <
\lambda_k ( {\bm C}_y ) ,
\end{equation}
where the last inequality is due to condition \eqref{eq:s_gap}. 
We can now apply the Davis-Kahan $\sin(\Theta)$ theorem \cite{daviskahan1970}:
\begin{equation}
{\left\| \sin\Theta\left({\bm V}_k, \widehat{\bm V}_k\right) \right\|_2} \leq \frac{\|\bm{\Delta}\|_2}{\lambda_{(2)}-\lambda_{(3)}-\|{\bm C}_y-\widehat{\bm C}_y^m\|_2}.
\end{equation}
Substituting the above into \eqref{eq:err-eigvec-bound} and then into \eqref{eq:l_chain} yields the desired bound \eqref{eq:d_bd}.
\end{IEEEproof}

{When the sample bound~\eqref{eq:convbd2} is attained, \eqref{eq:s_gap} holds.
Then, combining} \Cref{lem:lowerbd} and \Cref{prop:exact} shows that if
\begin{equation} \label{eq:gap_bd}
\| {\bm C}_y - \widehat{\bm C}_y^m \|_2 \leq \frac{\frac{n}{k}(c_1-c_2)}{1 + \sqrt{ 2(n+k)}},
\end{equation}
then \Cref{alg:timevary} always returns a partition with zero error rate, relative to the true PPM communities.
Applying \Cref{prop:conc} with the boundedness assumption on $\|{\bm y}^\tl\|_2$ {shows that the sampling condition fulfills~\eqref{eq:gap_bd}, as desired}.

\begin{Remark}
  Due to the non-convexity of the $k$-means objective, the assumption made in \Cref{prop:exact} that Algorithm~\ref{alg:timevary} yields an optimal solution is not practical.
  However, under certain conditions, there are algorithms that guarantee $(1+\epsilon)$ optimality~\cite{Kumar04simple}, \ie that bound the suboptimality gap with respect to the true non-convex optimum.
  Under these conditions, an expression analogous to \eqref{eq:d_bd} can be derived.
  More precisely, we obtain
  \begin{equation}\label{eq:d_bd_epsilon}
    F(g,{\bm V}_k) \leq \frac{k (2 + \epsilon)^2}{\nu^2} \|{\bm C}_y - \widehat{\bm C}_y^m\|_2^2 \eqs.
  \end{equation}
  Ultimately, the required sampling rate in \Cref{thm:2} remains unchanged even if one can only guarantee a near-optimal solution of the $k$-means problem.
\end{Remark} 

\section{Numerical Experiments and Applications}\label{sec:numerical_exp}

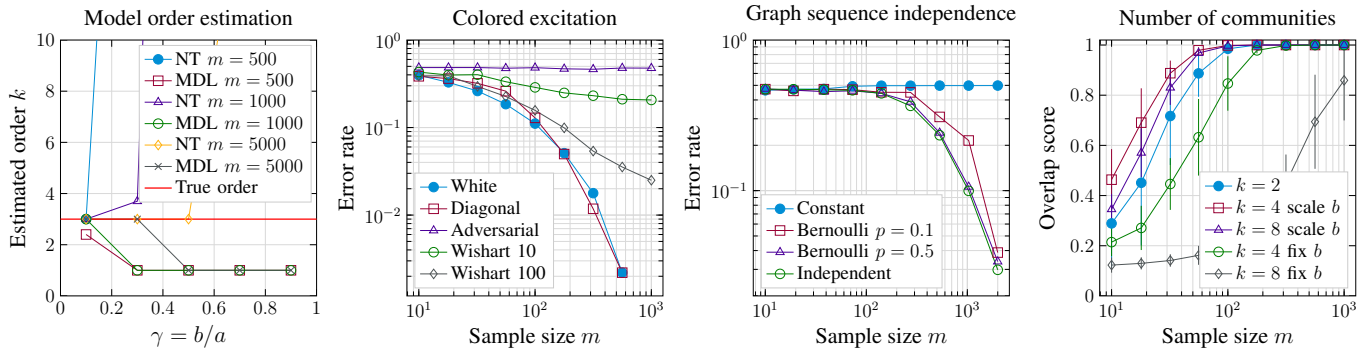
\begin{figure*}[t]
  \centering
  \resizebox{\linewidth}{!}{\input{num_experiments/combined-partition-recovery.tex}}
  \vspace{-0.5cm}
  \caption{Numerical experiments on synthetic graphs. (Left) Model order selection task for PPM with $k=3$ communities averaged over 10 experiments.
    (Center-left) Error rate of community detection ($k=2$) for colored excitation averaged over 10 experiments.
    (Center-right) Error rate of community detection ($k=2$) for dependent graph sequences averaged over 10 experiments.
    (Right) Overlap score of community detection for varying number of communities averaged over 20 experiments.}
  \label{fig:exp_comdet}
\end{figure*}

\begin{figure*}[t]
  \centering
  \includegraphics[width=0.8\linewidth]{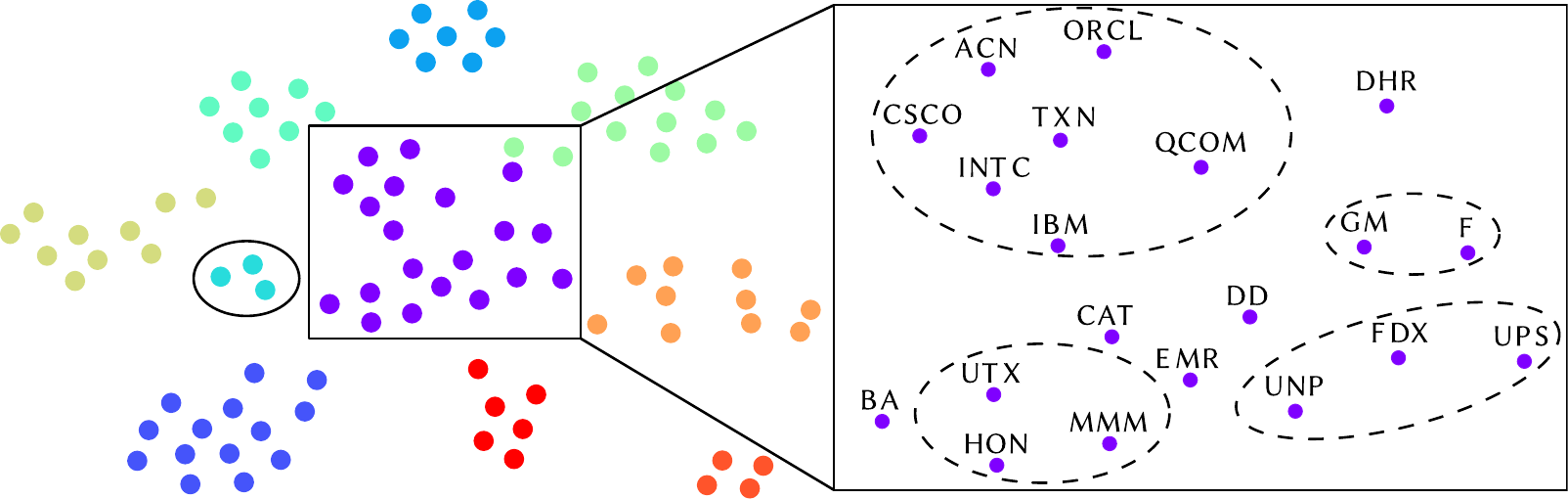}
  \caption{t-SNE embedding of the top eigenvectors for the stock dataset.
    The largest detected sector (boxed) is shown in detail, containing automotive companies (Ford, General Motors), technology and technology consulting companies (Cisco, Intel, IBM, Oracle), and conglomerates (3M, Honeywell, United Technologies), and others.
    Although the $k$-means clustering algorithm grouped these distinct sectors into one large community, this embedding demonstrates how the top eigenvectors of $\widehat{\bm C}_y^m$ capture the community structure.
  We note that the smaller communities, such as the community of three companies (circled on the left) containing defense contractors (General Dynamics, Lockheed Martin, Raytheon), exhibit stronger coherence in terms of market sector.}
  \label{fig:exp_stock_embedding}
\end{figure*}

\begin{figure*}[t]
  \centering
  \resizebox{\linewidth}{!}{\input{num_experiments/combined-stock-plot.tex}}
  \vspace{-0.5cm}
  \caption{Numerical experiments on stock dataset. (Left) Pairwise labeling success rates for non-overlapping, contiguous sets of samples from the stock dataset, as well as a random labeling for comparison.
    (Center-left) Covariance matrix $\widehat{\bm C}_y^M$ for the stock dataset. The partitions are outlined in red, matching with the blocks of higher covariance along the diagonal of $\widehat{\bm C}_y^M$.
    (Center-right) Order selection task for the stock dataset, averaged over 50 runs.
    (Right) Error rate of community detection task for the stock dataset, for both the full pipeline and fixed $k=10$. As $m$ increases, the full pipeline better predicts $k$, causing the two lines to merge. Each line is the average error rate over 50 runs.}
  \label{fig:exp_stock}
\end{figure*}
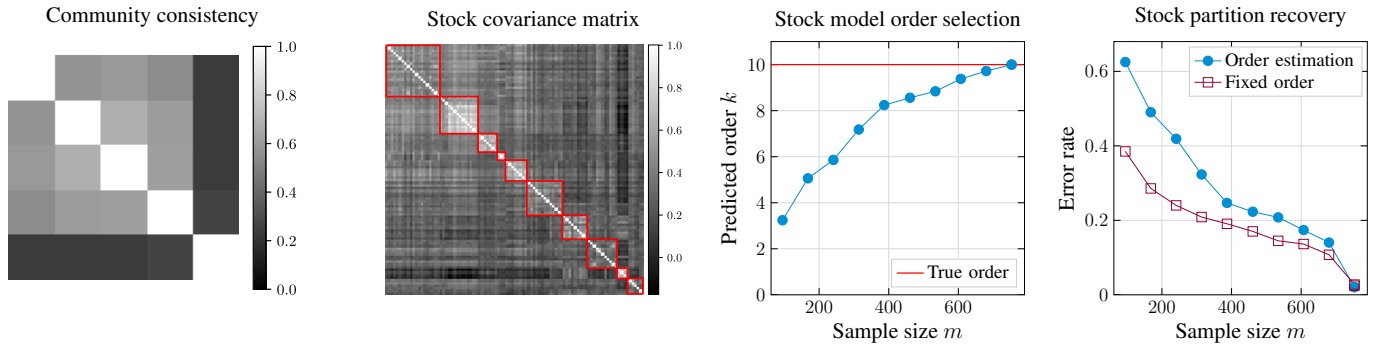

In this section, we demonstrate the performance of the proposed methods for model order selection and partition recovery in both synthetic and real-world data.
Unless specified otherwise, experiments are run with graphs of size $n=500$ nodes, and PPM model parameters $a=4\log{n},  b=\gamma a$ where $\gamma=0.3$.
As done in Section~\ref{ssec:toy_example}, we consider a network process represented by the graph filter $\mathcal{H}({\bm L}^{(\ell)})=({\bm I}-\beta{\bm L}^{(\ell)})^5$, with $\beta=1/(4+4\gamma)\log{n}$.
To benchmark the performance of community detection, we define the permutation-invariant error rate of a predicted labeling $\widehat{g}$ on $\mathcal{N}$:
\begin{equation}\label{eq:errorrate}
  r_{err}=\frac{1}{|{\cal N}|}\min_{\sigma\in {\cal S}_k}|\{x\in {\cal N}\colon g_x\neq \sigma(\widehat{g}_x)\}|.
\end{equation}
where ${\cal S}_k$ is the set of permutations $\sigma\colon[1,\ldots,k]\to[1,\ldots,k]$.

\subsection{Model order selection}
\label{ssec:exp_order_selection}

We consider a PPM with $k=3$ communities and analyze the estimated order $k^\star$ as a function of the ratio $\gamma = b/a$ for different number of observed signals $m$; see Fig.~\ref{fig:exp_comdet}~(left).
We focus on two different methods to obtain $k^\star$:
\begin{enumerate*}[label=\inlinelabel]
\item the MDL method described in \eqref{eq:mdl} and
\item a naive thresholding method that counts the number of eigenvalues greater than the mid-point $\delta_{\sf th} = (\lambda_{(2)} + \lambda_{(3)})/2$ between the $k$th and the $(k+1)$th eigenvalue of the \emph{true} covariance $\bm C_y$.
\end{enumerate*}

We call this method naive, since it does not take into account the noise stemming from considering a finite number of observations $m$.
Note that while this method would be optimal for $m \to \infty$, it requires knowledge of $\bm C_y$, and is thus not implementable in practice.

The results in Fig.~\ref{fig:exp_comdet}~(left) indicate that both estimators perform well when $\gamma$ is small, \ie when the parameters of the PPM yield an easily detectable model.
Naturally, estimators based on larger number of observed signals $m$ are more robust to increasing values of $\gamma$, with both methods estimating the correct order for $\gamma = 0.3$ when $m = 5000$.
However, when $\gamma$ grows large enough, both estimators behave in distinct ways:
the MDL method tends to underestimate the model order whereas the naive threshold method overestimates it.

To see why this is the case, notice that the first term in the MDL expression \eqref{eq:mdl} promotes orders $p$ for which the lower $n-p$ eigenvalues of $\widehat{\bm C}_y^m$ are flat whereas the second term penalizes high model orders. 
Thus, whenever there is no clear jump between the $k$th and the $(k+1)$th eigenvalues because $\gamma$ is large, the second term in \eqref{eq:mdl} dominates and the model order is reduced to its minimum of $1$.
In contrast, the naive threshold method simply counts the number of eigenvalues of $\widehat{\bm C}_y^m$ greater than $\delta_{\sf th} = (\lambda_{(2)} + \lambda_{(3)})/2$.
For small $m$, the eigenvalues of $\widehat{\bm C}_y^m$ are strongly perturbed versions of those of ${\bm C}_y$.
Thus, some of these perturbed eigenvalues tend to exceed the threshold $\delta_{\sf th}$, increasing the estimated order $k^\star$.
Moreover, as $\gamma$ increases, the gap between $\lambda_{(2)}$ and $\lambda_{(3)}$ becomes smaller, rendering it more likely for eigenvalues to cross the threshold by chance.

\subsection{Colored excitation}
\label{sec:exp_colored}

For a PPM of $k$ communities, our algorithm is predicated upon the system $\mathcal{H}({\bm L}^{(\ell)})$ being approximately rank-$k$ (plus a multiple of the identity).
If the excitation ${\bm w}^{(\ell)}$ is white ($\mathbb{E}[{\bm w}^{(\ell)}({\bm w}^{(\ell)})^\top]={\bm I}$, $\mathbb{E}[{\bm w}^{(\ell)}]={\bm 0}$), the eigenvectors of the system are excited uniformly.
Accordingly, the covariance matrix $\widehat{\bm C}_y^m$ will be
\begin{enumerate*}[label=\inlinelabel]
\item approximately described by a rank-$k$ matrix plus a multiple of the identity and 
\item have $k$ top eigenvectors that (approximately) capture the community structure.
\end{enumerate*}

We relax the assumption of a white excitation by coloring the excitation signal and observing the partition recovery performance of our algorithm as a function of the number of samples $m$.

We consider the following different scenarios for our excitation signal.
The white excitation is drawn from the normal distribution ${\cal N}({\bm 0},{\bm I}_n)$.
The diagonal excitation varies the diagonal entries of the covariance matrix, drawing the excitation from the distribution ${\cal N}({\bm 0},\text{diag}({\cal U}[0,1]^n))$.
The Wishart excitation is drawn from a Gaussian distribution whose covariance matrix is a Wishart matrix of $p$ samples.
That is, ${\bm w}^{(\ell)}\sim{\cal N}({\bm 0},{\bm W}_n({\bm I},p))$.
Clearly, as $p\to\infty$, ${\bm W}_n({\bm I},p)\to{\bm I}$, approaching the white excitation.
So, we consider the case where $p<n$, yielding a rank-deficient covariance matrix.
Finally, the adversarial excitation colors the excitation covariance to be strongly biased towards the lower $n-k$ eigenvectors of the system.
Specifically, for $\EE[{\bm A}]=[{\bm V}_1 {\bm V}_2]{\bm \Lambda}[{\bm V}_1 {\bm V}_2]^\top$, where the top $k$ and lower $n-k$ eigenvector matrices are denoted by ${\bm V}_1$ and ${\bm V}_2$, respectively, the excitation is drawn from the distribution ${\cal N}({\bm 0},0.01{\bm V}_1{\bm V}_1^\top+0.81{\bm V}_2{\bm V}_2^\top)$.

The only case that preserves both the identically distributed and independence conditions at each node is the white excitation.
The diagonal excitation maintains independence at each node (since the covariance matrix is diagonal), but the variances are different across nodes, breaking the identically distributed condition.
The Wishart and adversarial cases break both conditions.
Note that the Wishart excitation has a chance of not exciting the leading eigenvectors of the system, resulting in poor performance, but could also excite all (or some) of them, due to the rank-deficient nature of its covariance matrix.
Moreover, the adversarial excitation very weakly excites the leading eigenvectors, almost guaranteeing poor performance.

The results in Fig.~\ref{fig:exp_comdet}~(center-left) indicate that the diagonal input closely matches the white input's performance, suggesting that having identically distributed input at each node is not very important, as long as independence is maintained.

When the independence condition is broken, as in the rank-deficient Wishart excitations, the performance of our algorithm degrades.
However, as the rank $p$ of the excitation increases, the average performance increases as well.
This is intuitive, as a higher-rank excitation is more likely to excite the top eigenvectors of the system, improving the algorithm's performance.

Finally, the adversarial excitation does no better than a random guess  (error rate $\approx0.5$) for all $m$.
This confirms our expectation, as this input scheme does not excite the top $k$ eigenvectors that reflect the community structure of the PPM.

\subsection{Graph sequence independence}
\label{sec:exp_sequence}

In the studied setting of only observing signals on a graph (as opposed to the graph itself), our algorithm is able to successfully detect communities in a PPM even though the parameters of the PPM are below known detectability thresholds \cite{mossel2014reconstruction}.
This non-intuitive result is due to the fact that each observation corresponds to an independent initial condition and an \emph{independent} realization of the underlying PPM.
We can leverage the information from these independent draws by averaging over many different samples, thereby sidestepping the detectability limit (which assumes that we observe a single graph).

To demonstrate how our algorithm leverages the implicit observation of many realizations of an
 undetectable PPM for a given number of samples $m$, we conduct the following experiment.
Instead of sampling an independent graph for each initial condition, we consider a sequence of graphs modeled by a Bernoulli (graph)-process:
starting with some realization of the PPM $\mathcal{G}^{(\ell)}$, let the next realization $\mathcal{G}^{(\ell+1)}=\mathcal{G}^{(\ell)}$ be the same graph with probability $1-p$, and draw $\mathcal{G}^{(\ell)}$ randomly from the PPM with probability $p$.
So, when $p=1$, this is equivalent to drawing $m$ graphs independently, and when $p=0$, this is equivalent to only using a single graph.
Note, however, that we still observe the system for $m$ independently drawn white excitation signals ${\bm w}^{(\ell)}\sim\mathcal{N}({\bm 0},{\bm I})$.

The results in Fig.~\ref{fig:exp_comdet}~(center-right) reflect our intuition: the partition recovery performance improves with the Bernoulli parameter $p$.

Specifically, the $p=0$ (constant) case fails for all $m$, since performance is upper bounded by the case when the graph is directly observed.
The considered PPM is undetectable when observing a single graph, so our algorithm fails.
The $p=0.1$ case does better than the constant case, but does not achieve the required number of graphs observed to match the $p=1$ (independent) case.
However, the $p=0.5$ case behaves quite similarly to the independent sequence, suggesting there may be some point at which it is sufficient to excite a few ($pm$) graphs multiple times each, rather than excite many ($m$) graphs once each.

\subsection{Signal-to-noise ratio from community structure}

To understand how the community structure influences our algorithm's performance, we measure the \emph{overlap score} between the predicted labeling and the true labeling for an increasing number of communities $k$, where all communities have a fixed size $n_c$.
That is, for some $k$, the PPM will have $n=kn_c$ nodes.
The overlap score, defined in \cite{Schaub2018} as
\begin{equation} \label{eq:overlap_score}
  Z=\frac{z_{actual}-z_{chance}}{1-z_{chance}}
\end{equation}
where $z_{actual}$ is the fraction of correctly labeled nodes, and $z_{chance}=1/k$ is the probability of correctly guessing a node's true group assignment.
So, an overlap score of $0$ indicates that the candidate labeling is no better than a random guess, and an overlap score of $1$ indicates a perfect match.
Using this metric rather than the error rate accounts for the different community structures, allowing for a fair comparison.

As $k$ increases, $a/(kn_c)$ is kept constant, but $b/(kn_c)$ is either
\begin{enumerate*}[label=\inlinelabel]
\item kept constant or
\item scaled to keep the expected number of inter-cluster edges incident to each community constant.
\end{enumerate*}

That is,
\begin{equation} \label{eq:constant_interference}
  \frac{b}{kn_c}(k-1)n_c=b\frac{k-1}{k}
\end{equation}
is constant for all $k$, where it is implicit that $b$ varies with $k$.
In fact, using the signal to noise ratio (SNR) defined in \cite{Abbe2018}, this scheme where $b/(kn_c)$ is scaled with $k$ yields
\begin{equation} \label{eq:communities_snr}
  \text{SNR}(k)\sim\left(1-\frac{\gamma_2}{k-1}\right)^2,
\end{equation}
where $\gamma_2$ is the ratio $b/a$ when $k=2$.

Fig.~\ref{fig:exp_comdet}~(right) illustrates the error rate for each graph model.
When $b/(kn_c)$ is fixed, the performance degrades with $k$, as the ratio of inter-cluster to intra-cluster edges increases linearly with $k$.
However, normalizing $b/(kn_c)$ maintains consistent performance for all values of $k$.

\subsection{Clustering stock data}
\label{sec:exp_stock}

We apply our algorithm to the task of inferring community structure in the S\&P 100 stock market index.
The daily closing prices for 92 stocks from 4 January 2016 to 30 December 2018 were obtained from Yahoo! Finance\footnote{\url{https://finance.yahoo.com}}, and the daily log-returns calculated.
These daily returns are then normalized to have zero mean and unit variance.

The assumption here is that stocks have an underlying community structure dictated by a stochastic block model, and that the log-returns for each day are independent and the result of a filter on a graph drawn from an SBM.
These are clearly very strong assumptions to make for real data, so we justify the application of our algorithm to this dataset by measuring its ``stability'' over the time-series.
We split the dataset into 4 contiguous blocks of size $m=\lfloor M/4\rfloor$, where $M$ is the total number of samples, and apply our algorithm to $\widehat{\bm C}_y^m$ with $k=10$ to recover communities for different periods of time.
A random labeling is also generated for reference.
Then, the pairwise success rate ($1-\text{error rate}$) between each sample set is computed [\lcf~\eqref{eq:errorrate}], as shown in Fig.~\ref{fig:exp_stock}~(left).
Although the detected communities do not overlap perfectly, they clearly exhibit a high degree of consistency.

Obviously, there is no ground-truth partition of the stock data, so we apply our algorithm to the whole dataset, yielding $k=10$ communities as determined by the MDL (Algorithm~\ref{alg:model_select}).
Fig.~\ref{fig:exp_stock_embedding} shows the t-SNE embedding \cite{maaten2008visualizing} of the dominant eigenvectors.
Despite companies often not strictly belonging to one sector, the communities seem to capture obvious commonalities between companies\footnote{For a table of detected communities, as well as code for the other experiments, see \url{https://github.com/tmrod/timevary-netfeat-supplement}}.
Moreover, the block structure of the covariance matrix for all $M$ samples, shown in Fig.~\ref{fig:exp_stock}~(center-left), is apparent.
This further justifies treating the dataset as if it is driven by an SBM.
Proceeding under the assumption that this ``final'' result reflects the true community structure, we evaluate our algorithm with respect to order selection and community detection.

Fig.~\ref{fig:exp_stock}~(center-right) shows the order estimation with respect to the sample size.
For some number of samples $m$, a single ``block'' of $m$ consecutive samples is observed, from which the model order is inferred.
This tends to underestimate the true model order as observed in Section~\ref{ssec:exp_order_selection}.

Note that in the full processing pipeline ($\widehat{\bm C}_y^m\to k^\star\to g$), an incorrect model order estimate $k^\star$ will lead to poor classification performance.
Hence, in Fig.~\ref{fig:exp_stock}~(right) we show the community detection performance for both the full pipeline (where $k^\star$ is determined via MDL) and for fixed $k^\star=10$ (both using the same consecutive sampling scheme as before).
Fig.~\ref{fig:exp_stock}~(right) shows an expected drop in the classification error rate for both settings as $m$ increases, with the gap between the two curves closing as the estimate $k^\star$ becomes more accurate.

\section{Discussion}
\label{sec:discussion}
In this work, we considered the `blind' community detection problem, where we observe signals on the nodes of a family of graphs, rather than the edges of the graph itself.
Stated differently, the observations correspond to the output of a (time-varying, random) graph filter applied to random initial conditions [\lcf~the system model \eqref{eq:signal_model}].
Assuming that the underlying graphs correspond to (unobserved) realizations of a PPM, we aim to infer the number of communities present as well as the corresponding partition of the nodes.
We propose the use of spectral algorithms on the empirical covariance of these signals to infer the latent structure of the graph sequence.
We show that our algorithms have statistical performance guarantees for both the asymptotic and finite sampling cases, as also demonstrated via extensive numerical experiments.

There are many potential avenues for future research.
{Our proof of the MDL criterion's performance in the finite sample regime only provides sampling requirements to guarantee that the order is not underestimated.
As shown empirically in Section~\ref{ssec:exp_order_selection}, the model order is typically \emph{not} overestimated.
We leave further analysis of this behavior for future work.}
As shown in Section~\ref{sec:exp_colored}, the clustering algorithm does not strictly require the initial condition ${\bm w}^{(\ell)}$ to be white.
If the observer can manipulate the covariance matrix of the inputs, it appears possible to get better performance by adaptively coloring the inputs to the system.
Furthermore, Section~\ref{sec:exp_sequence} shows that the graph sequence does not need to be strictly independent for the algorithm to perform well.
Analysing these dependencies in more detail could yield algorithms with refined performance guarantees, \eg for more realistic sampling regimes in which there is a correlation over time in the observed graph signal --- a scenario that is highly relevant for data emerging from real-world applications.

\appendices

\section{Proof of \eqref{eq:taylor}} \label{app:small}
Observe that \eqref{eq:taylor} is equivalent to the following inequality
\begin{equation} \label{eq:transformed}
\gamma^x < 1 + x (\gamma - 1),
\end{equation}
where $x=(k-p)/(n-p) \in [0,1)$ and $\gamma = \lambda_{(2)}/\lambda_{(3)} > 1$. For any $\gamma > 1$, the truncated Taylor series of $\gamma^x$ gives 
\begin{equation}
\gamma^x = 1 + x (\gamma - 1) + \frac{ x(x-1) \xi^{x-2} }{2} (\gamma - 1)^2,
\end{equation}
where $\xi \in [1, \gamma]$. Finally, the last term is negative as $0 \leq x < 1$, we get \eqref{eq:transformed}. 

\bibliographystyle{IEEEtran}
\bibliography{extracted}

\end{document}

%% file: example_experiments/ResLim_T1.tikz.tex
\begin{tikzpicture}

  \begin{groupplot}[
    group style={group size=2 by 2,
    horizontal sep=2cm},
    every tick label/.append style={font=\large},
    legend style={font=\large,draw=white!80.0!black},
    log basis x={10},
    log basis y={10},
    tick align=inside,
    tick pos=both,
    xlabel={\Large Sample size \(\displaystyle m\)},
    x grid style={gray!30},
    xmajorgrids,
    xminorgrids,
    xmode=log,
    xtick style={color=black},
    xmin=10, xmax=3e5,
    ylabel={\Large Error rate},
    y grid style={gray!30},
    ymajorgrids,
    yminorgrids,
    ymode=log,
    ytick style={color=black},
    ymin=1e-4,
    width=0.4\textwidth,
    height=0.4\textwidth
  ]

  \nextgroupplot[
    title={\Large Covariance $m=10$},
    xmin=0,xmax=1,ymin=0,ymax=1,
    xmode=normal,ymode=normal,
    xlabel={},ylabel={},
    axis lines=none,
    axis background/.style={fill=white},
    width=0.44\textwidth,
    ]
  \addplot graphics [xmin=0,xmax=1,ymin=0,ymax=1] {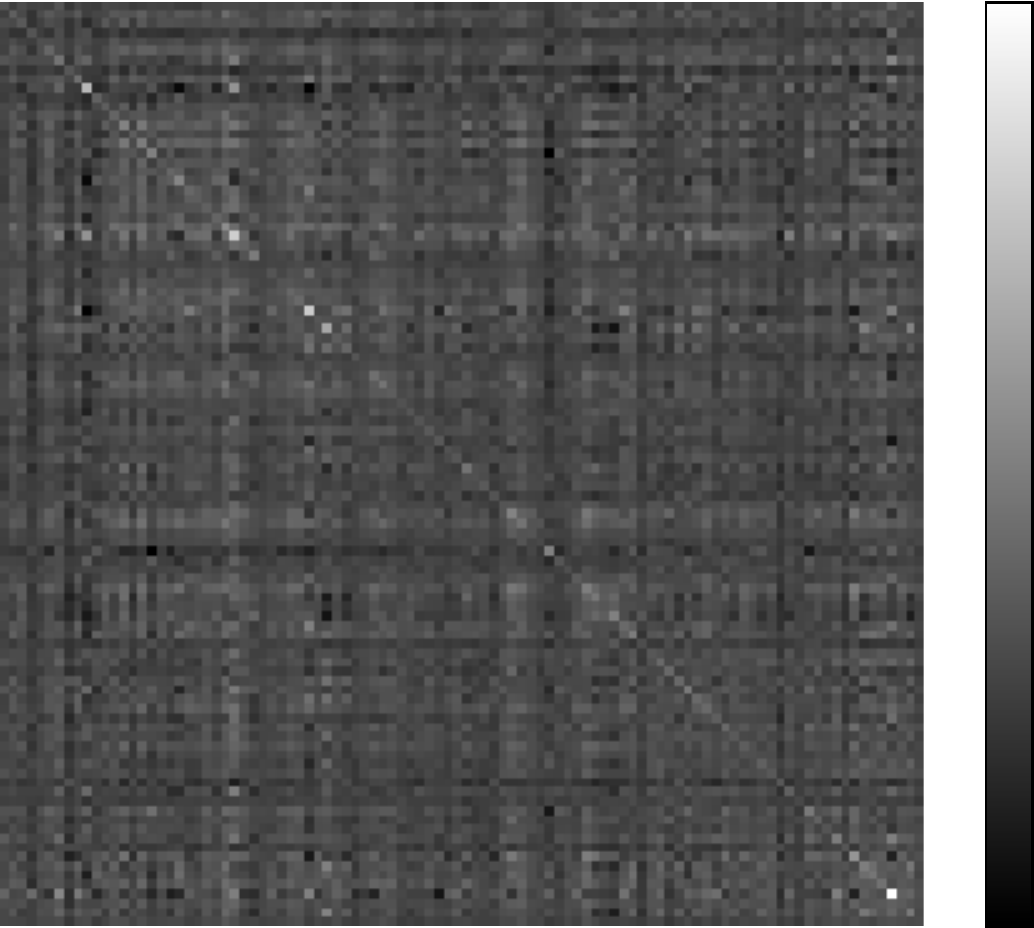};

  \nextgroupplot[
    title={\Large Covariance $m=10^4$},
    xmin=0,xmax=1,ymin=0,ymax=1,
    xmode=normal,ymode=normal,
    xlabel={},ylabel={},
    axis lines=none,
    width=0.44\textwidth,
    ]
  \addplot graphics [xmin=0,xmax=1,ymin=0,ymax=1] {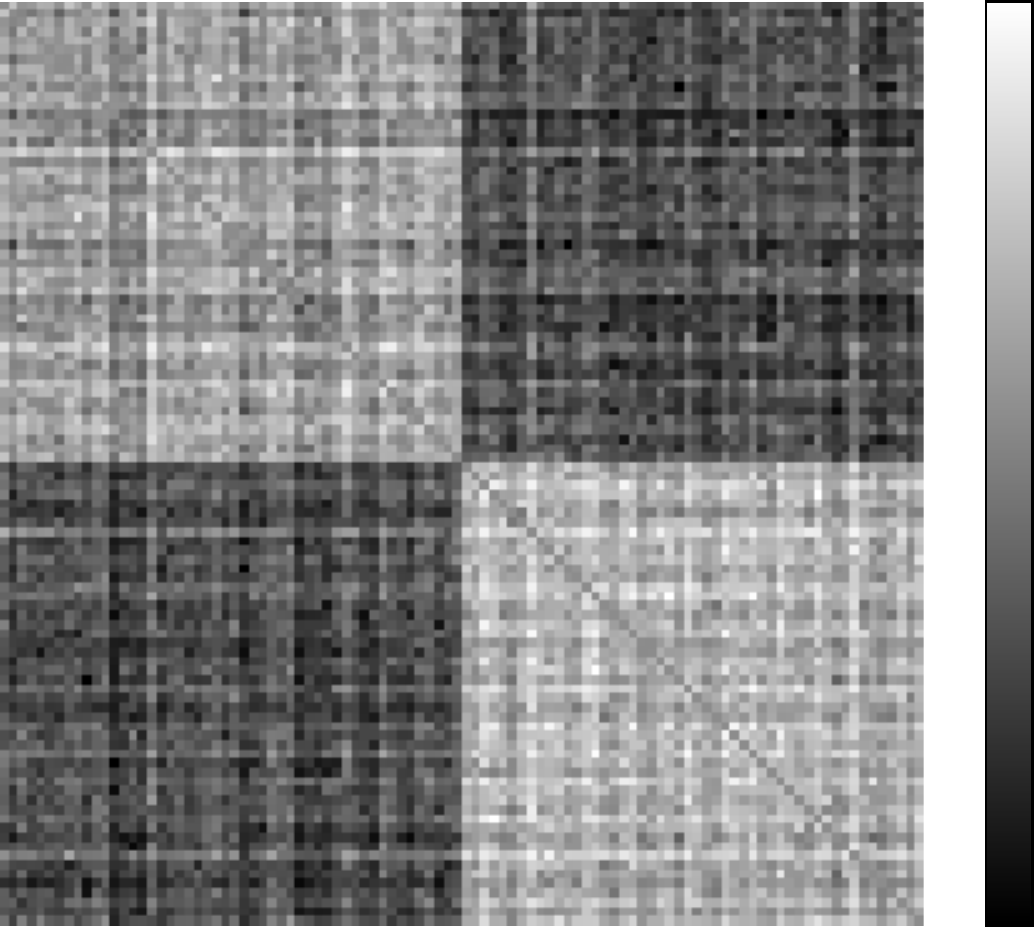};

  \nextgroupplot[
    title={\Large Model order estimation},
    legend cell align=left,
    legend style={legend pos=south west}
  ]
  \addplot[asublue, very thick, mark options={solid,mark size=3}, mark=*] 
  table[x index=0, y index=1, col sep=comma] {./example_experiments/ResLim_01_err_Kest.csv};
  \addlegendentry{$\gamma = 0.1$};
  \addplot[asured, very thick, mark options={solid,mark size=3}, mark=square] 
  table[x index=0, y index=1, col sep=comma] {./example_experiments/ResLim_03_err_Kest.csv};
  \addlegendentry{$\gamma = 0.3$};
  \addplot[asured!50!blue, very thick, mark options={solid,mark size=3}, mark=triangle] 
  table[x index=0, y index=1, col sep=comma] {./example_experiments/ResLim_05_err_Kest.csv};
  \addlegendentry{$\gamma = 0.5$};
  \addplot[green!50!black, very thick, mark options={solid,mark size=3}, mark=o] 
  table[x index=0, y index=1, col sep=comma] {./example_experiments/ResLim_07_err_Kest.csv};
  \addlegendentry{$\gamma = 0.7$};
  \addplot[asuorange, very thick, mark options={solid,mark size=3}, mark=diamond] 
  table[x index=0, y index=1, col sep=comma] {./example_experiments/ResLim_09_err_Kest.csv};
  \addlegendentry{$\gamma = 0.9$};

  \nextgroupplot[
    title={\Large Partition recovery},
    legend cell align=left,
    legend style={legend pos=south west},
  ]
  \addplot[asublue, very thick, mark options={solid,mark size=3}, mark=*] 
  table[x index=0, y index=1, col sep=comma] {./example_experiments/ResLim_01a_err_CYnl.csv};
  \addlegendentry{$\gamma = 0.1$};
  \addplot[asured, very thick, mark options={solid,mark size=3}, mark=square] 
  table[x index=0, y index=1, col sep=comma] {./example_experiments/ResLim_03a_err_CYnl.csv};
  \addlegendentry{$\gamma = 0.3$};
  \addplot[asured!50!blue, very thick, mark options={solid,mark size=3}, mark=triangle] 
  table[x index=0, y index=1, col sep=comma] {./example_experiments/ResLim_05a_err_CYnl.csv};
  \addlegendentry{$\gamma = 0.5$};
  \addplot[green!50!black, very thick, mark options={solid,mark size=3}, mark=o] 
  table[x index=0, y index=1, col sep=comma] {./example_experiments/ResLim_07a_err_CYnl.csv};
  \addlegendentry{$\gamma = 0.7$};
  \addplot[asuorange, very thick, mark options={solid,mark size=3}, mark=diamond] 
  table[x index=0, y index=1, col sep=comma] {./example_experiments/ResLim_09a_err_CYnl.csv};
  \addlegendentry{$\gamma = 0.9$};

  \end{groupplot}
\end{tikzpicture}

%% file: num_experiments/combined-partition-recovery.tex
\begin{tikzpicture}
  \begin{groupplot}[
    group style={group size=4 by 1,
      horizontal sep=2cm},
    every tick label/.append style={font=\large},
    legend style={font=\large},
    log basis x={10},
    log basis y={10},
    tick align=inside,
    tick pos=both,
    xlabel={\Large Sample size \(\displaystyle m\)},
    x grid style={gray!30},
    xmajorgrids,
    xminorgrids,
    xlabel near ticks,
    ylabel near ticks,
    xmode=log,
    xtick style={color=black},
    y grid style={gray!30},
    ymajorgrids,
    yminorgrids,
    ytick style={color=black},
    width=0.4\textwidth,
    height=0.4\textwidth
    ]

    \nextgroupplot[
      title={\Large Model order estimation},
      legend cell align={left},
      legend style={draw=white!80.0!black},
      xmin=0.00, xmax=1.00,
      xmode=normal,
      xlabel={\Large $\gamma=b/a$},
      ylabel={\Large Estimated order \(\displaystyle k\)},
      ymin=0, ymax=10
    ]
    \addplot [semithick, asublue, mark=*, mark size=3, mark options={solid}]
    table {%
      0.1 3
      0.3 36.5
      0.5 91.8
      0.7 123.2
      0.9 136.9
    };
    \addlegendentry{NT $m=500$}
    \addplot [semithick, asured, mark=square, mark size=3, mark options={solid,fill opacity=0}]
    table {%
      0.1 2.4
      0.3 1
      0.5 1
      0.7 1
      0.9 1
    };
    \addlegendentry{MDL $m=500$}
    \addplot [semithick, asured!50!blue, mark=triangle, mark size=3, mark options={solid,fill opacity=0}]
    table {%
      0.1 3
      0.3 3.7
      0.5 61.6
      0.7 105.6
      0.9 126.1
    };
    \addlegendentry{NT $m=1000$}
    \addplot [semithick, green!50!black, mark=*, mark size=3, mark options={solid,fill opacity=0}]
    table {%
      0.1 3
      0.3 1
      0.5 1
      0.7 1
      0.9 1
    };
    \addlegendentry{MDL $m=1000$}
    \addplot [semithick, asuorange, mark=diamond, mark size=3, mark options={solid,fill opacity=0}]
    table {%
      0.1 3
      0.3 3
      0.5 3
      0.7 15.3
      0.9 46.4
    };
    \addlegendentry{NT $m=5000$}
    \addplot [semithick, asugrey, mark=x, mark size=3, mark options={solid}]
    table {%
      0.1 3
      0.3 3
      0.5 1
      0.7 1
      0.9 1
    };
    \addlegendentry{MDL $m=5000$}
    \addplot [thick, red]
    table {%
      0.00 3
      1.00 3
    };
    \addlegendentry{True order}
  
    \nextgroupplot[
      title={\Large Colored excitation},
      legend cell align={left},
      legend style={at={(0.03,0.03)}, anchor=south west, draw=white!80.0!black},
      xmin=7.9, xmax=1259,
      ymax=1.0,
      ylabel={\Large Error rate},
      ymode=log,
    ]
    \addplot [semithick, asublue, mark=*, mark size=3, mark options={solid}]
    table {%
      10 0.3958
      18 0.3282
      32 0.2614
      56 0.185
      100 0.111
      178 0.051
      316 0.0178
      562 0.0022
      1000 0
    };
    \addlegendentry{White}
    \addplot [semithick, asured, mark=square, mark size=3, mark options={solid,fill opacity=0}]
    table {%
      10 0.3852
      18 0.3626
      32 0.32
      56 0.2604
      100 0.129
      178 0.0498
      316 0.0118
      562 0.0022
      1000 0
    };
    \addlegendentry{Diagonal}
    \addplot [semithick, asured!50!blue, mark=triangle, mark size=3, mark options={solid,fill opacity=0}]
    table {%
      10 0.4842
      18 0.4844
      32 0.4846
      56 0.4762
      100 0.4828
      178 0.4686
      316 0.4634
      562 0.4788
      1000 0.4772
    };
    \addlegendentry{Adversarial}
    \addplot [semithick, green!50!black, mark=*, mark size=3, mark options={solid,fill opacity=0}]
    table {%
      10 0.43
      18 0.3982
      32 0.4016
      56 0.3338
      100 0.2876
      178 0.2488
      316 0.2312
      562 0.2106
      1000 0.2066
    };
    \addlegendentry{Wishart 10}
    \addplot [semithick, asugrey, mark=diamond, mark size=3, mark options={solid,fill opacity=0}]
    table {%
      10 0.3896
      18 0.39
      32 0.2954
      56 0.2264
      100 0.1572
      178 0.0992
      316 0.0538
      562 0.0354
      1000 0.025
    };
    \addlegendentry{Wishart 100}

    \nextgroupplot[
      title={\Large Graph sequence independence},
      legend cell align={left},
      legend style={at={(0.03,0.03)}, anchor=south west, draw=white!80.0!black},
      xmin=7.6, xmax=2607,
      ylabel={\Large Error rate},
      ymin=0.02, ymax=1.0,
      ymode=log
    ]
    \addplot [semithick, asublue, mark=*, mark size=3, mark options={solid}]
    table {%
      10 0.4632
      19 0.4706
      38 0.4736
      73 0.493
      141 0.4974
      274 0.4974
      532 0.498
      1031 0.498
      2000 0.4984
    };
    \addlegendentry{Constant}
    \addplot [semithick, asured, mark=square, mark size=3, mark options={solid,fill opacity=0}]
    table {%
      10 0.4718
      19 0.4594
      38 0.4716
      73 0.4682
      141 0.4498
      274 0.449
      532 0.3086
      1031 0.2152
      2000 0.0388
    };
    \addlegendentry{Bernoulli $p=0.1$}
    \addplot [semithick, asured!50!blue, mark=triangle, mark size=3, mark options={solid,fill opacity=0}]
    table {%
      10 0.4738
      19 0.4644
      38 0.4542
      73 0.46
      141 0.4426
      274 0.3886
      532 0.242
      1031 0.106
      2000 0.0338
    };
    \addlegendentry{Bernoulli $p=0.5$}
    \addplot [semithick, green!50!black, mark=*, mark size=3, mark options={solid,fill opacity=0}]
    table {%
      10 0.471
      19 0.4712
      38 0.4676
      73 0.4636
      141 0.4424
      274 0.3654
      532 0.2332
      1031 0.0996
      2000 0.0298
    };
    \addlegendentry{Independent}
  
    \nextgroupplot[
      title={\Large Number of communities},
      legend cell align={left},
      legend style={at={(0.97,0.03)}, anchor=south east, draw=white!80.0!black},
      xmin=7.9, xmax=1259.0,
      ylabel={\Large Overlap score},
      ymin=0.00, ymax=1.02
    ]
    \path [draw=asublue, semithick]
    (axis cs:10,0.112877883274133)
    --(axis cs:10,0.465122116725867);

    \path [draw=asublue, semithick]
    (axis cs:18,0.20172304558985)
    --(axis cs:18,0.70027695441015);

    \path [draw=asublue, semithick]
    (axis cs:32,0.521830842600579)
    --(axis cs:32,0.912169157399421);

    \path [draw=asublue, semithick]
    (axis cs:56,0.793560179794694)
    --(axis cs:56,0.980439820205307);

    \path [draw=asublue, semithick]
    (axis cs:100,0.968416876048223)
    --(axis cs:100,1.00158312395178);

    \path [draw=asublue, semithick]
    (axis cs:178,1)
    --(axis cs:178,1);

    \path [draw=asublue, semithick]
    (axis cs:316,1)
    --(axis cs:316,1);

    \path [draw=asublue, semithick]
    (axis cs:562,1)
    --(axis cs:562,1);

    \path [draw=asublue, semithick]
    (axis cs:1000,1)
    --(axis cs:1000,1);

    \path [draw=asured, semithick]
    (axis cs:10,0.341595910985068)
    --(axis cs:10,0.585070755681599);

    \path [draw=asured, semithick]
    (axis cs:18,0.553871605027407)
    --(axis cs:18,0.827461728305927);

    \path [draw=asured, semithick]
    (axis cs:32,0.837499169442091)
    --(axis cs:32,0.937834163891242);

    \path [draw=asured, semithick]
    (axis cs:56,0.966657660945618)
    --(axis cs:56,0.991342339054383);

    \path [draw=asured, semithick]
    (axis cs:100,0.995446581987385)
    --(axis cs:100,1.00122008467928);

    \path [draw=asured, semithick]
    (axis cs:178,1)
    --(axis cs:178,1);

    \path [draw=asured, semithick]
    (axis cs:316,1)
    --(axis cs:316,1);

    \path [draw=asured, semithick]
    (axis cs:562,1)
    --(axis cs:562,1);

    \path [draw=asured, semithick]
    (axis cs:1000,1)
    --(axis cs:1000,1);

    \path [draw=asured!50!blue, semithick]
    (axis cs:10,0.284620860459171)
    --(axis cs:10,0.405950568112258);

    \path [draw=asured!50!blue, semithick]
    (axis cs:18,0.473595978669686)
    --(axis cs:18,0.667261164187457);

    \path [draw=asured!50!blue, semithick]
    (axis cs:32,0.761271014343381)
    --(axis cs:32,0.898728985656619);

    \path [draw=asured!50!blue, semithick]
    (axis cs:56,0.955911248708233)
    --(axis cs:56,0.980374465577481);

    \path [draw=asured!50!blue, semithick]
    (axis cs:100,0.993680462774058)
    --(axis cs:100,0.999748108654514);

    \path [draw=asured!50!blue, semithick]
    (axis cs:178,1)
    --(axis cs:178,1);

    \path [draw=asured!50!blue, semithick]
    (axis cs:316,1)
    --(axis cs:316,1);

    \path [draw=asured!50!blue, semithick]
    (axis cs:562,1)
    --(axis cs:562,1);

    \path [draw=asured!50!blue, semithick]
    (axis cs:1000,1)
    --(axis cs:1000,1);

    \path [draw=green!50!black, semithick]
    (axis cs:10,0.158306554022008)
    --(axis cs:10,0.270360112644659);

    \path [draw=green!50!black, semithick]
    (axis cs:18,0.182851766262102)
    --(axis cs:18,0.359148233737898);

    \path [draw=green!50!black, semithick]
    (axis cs:32,0.342894875652727)
    --(axis cs:32,0.549105124347273);

    \path [draw=green!50!black, semithick]
    (axis cs:56,0.479605907287234)
    --(axis cs:56,0.785060759379432);

    \path [draw=green!50!black, semithick]
    (axis cs:100,0.738182156763691)
    --(axis cs:100,0.955151176569643);

    \path [draw=green!50!black, semithick]
    (axis cs:178,0.966815740776913)
    --(axis cs:178,0.990517592556421);

    \path [draw=green!50!black, semithick]
    (axis cs:316,0.994944949536696)
    --(axis cs:316,1.0010550504633);

    \path [draw=green!50!black, semithick]
    (axis cs:562,1)
    --(axis cs:562,1);

    \path [draw=green!50!black, semithick]
    (axis cs:1000,1)
    --(axis cs:1000,1);

    \path [draw=asugrey, semithick]
    (axis cs:10,0.0926199850164047)
    --(axis cs:10,0.153094300697881);

    \path [draw=asugrey, semithick]
    (axis cs:18,0.10548072827584)
    --(axis cs:18,0.153947843152731);

    \path [draw=asugrey, semithick]
    (axis cs:32,0.116261095934769)
    --(axis cs:32,0.166024618350945);

    \path [draw=asugrey, semithick]
    (axis cs:56,0.123967546430864)
    --(axis cs:56,0.199461024997707);

    \path [draw=asugrey, semithick]
    (axis cs:100,0.155677015146856)
    --(axis cs:100,0.266322984853144);

    \path [draw=asugrey, semithick]
    (axis cs:178,0.186475592549301)
    --(axis cs:178,0.376095836022127);

    \path [draw=asugrey, semithick]
    (axis cs:316,0.335271143429571)
    --(axis cs:316,0.563585999427572);

    \path [draw=asugrey, semithick]
    (axis cs:562,0.506627470019125)
    --(axis cs:562,0.88165824426659);

    \path [draw=asugrey, semithick]
    (axis cs:1000,0.699326854778984)
    --(axis cs:1000,1.01924457379244);

    \addplot [semithick, asublue, mark=*, mark size=3, mark options={solid}]
    table {%
      10 0.289
      18 0.451
      32 0.717
      56 0.887
      100 0.985
      178 1
      316 1
      562 1
      1000 1
    };
    \addlegendentry{$k=2$}
    \addplot [semithick, asured, mark=square*, mark size=3, mark options={solid, fill opacity=0}]
    table {%
      10 0.463333333333333
      18 0.690666666666667
      32 0.887666666666667
      56 0.979
      100 0.998333333333333
      178 1
      316 1
      562 1
      1000 1
    };
    \addlegendentry{$k=4$ scale $b$}
    \addplot [semithick, asured!50!blue, mark=triangle*, mark size=3, mark options={solid, fill opacity=0}]
    table {%
      10 0.345285714285714
      18 0.570428571428572
      32 0.83
      56 0.968142857142857
      100 0.996714285714286
      178 1
      316 1
      562 1
      1000 1
    };
    \addlegendentry{$k=8$ scale $b$}
    \addplot [semithick, green!50!black, mark=*, mark size=3, mark options={solid, fill opacity=0}]
    table {%
      10 0.214333333333333
      18 0.271
      32 0.446
      56 0.632333333333333
      100 0.846666666666667
      178 0.978666666666667
      316 0.998
      562 1
      1000 1
    };
    \addlegendentry{$k=4$ fix $b$}
    \addplot [semithick, asugrey, mark=diamond*, mark size=3, mark options={solid, fill opacity=0}]
    table {%
      10 0.122857142857143
      18 0.129714285714286
      32 0.141142857142857
      56 0.161714285714286
      100 0.211
      178 0.281285714285714
      316 0.449428571428571
      562 0.694142857142857
      1000 0.859285714285714
    };
    \addlegendentry{$k=8$ fix $b$}
  \end{groupplot}
\end{tikzpicture}

%% file: num_experiments/combined-stock-plot.tex
\begin{tikzpicture}
  \begin{groupplot}[
    group style={group size=4 by 1,
      horizontal sep=2cm},
    every tick label/.append style={font=\large},
    legend style={draw=white!80.0!black,font=\large},
    tick align=inside,
    tick pos=both,
    xlabel={\Large Sample size \(\displaystyle m\)},
    x grid style={gray!30},
    xlabel near ticks,
    ylabel near ticks,
    xmajorgrids,
    xminorgrids,
    xtick style={color=black},
    y grid style={gray!30},
    ymajorgrids,
    yminorgrids,
    ytick style={color=black},
    width=0.4\textwidth,
    height=0.4\textwidth,
    ]

    \nextgroupplot[
      title={\Large Community consistency},
      xmin=0,xmax=1,ymin=0,ymax=1,
      xlabel={},
      axis lines=none,
      width=0.4416\textwidth]
      \addplot graphics [xmin=0,xmax=1,ymin=0,ymax=1] {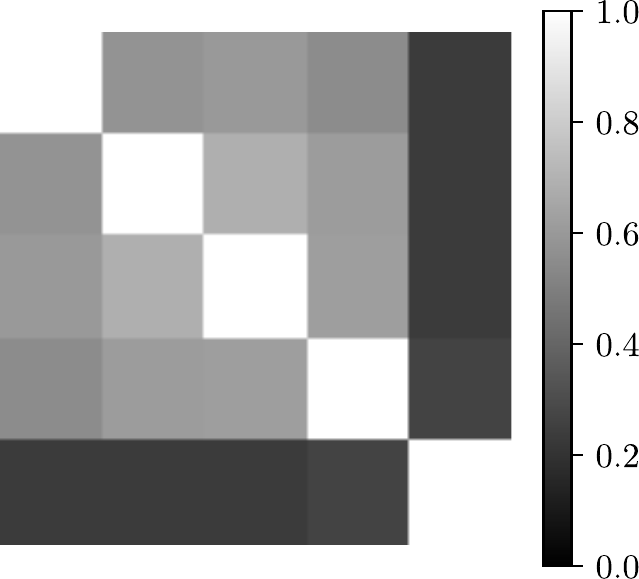};

    \nextgroupplot[
      title={\Large Stock covariance matrix},
      xmin=0,xmax=1,ymin=0,ymax=1,
      xlabel={},
      axis lines=none,
      width=0.4528\textwidth]
      \addplot graphics [xmin=0,xmax=1,ymin=0,ymax=1] {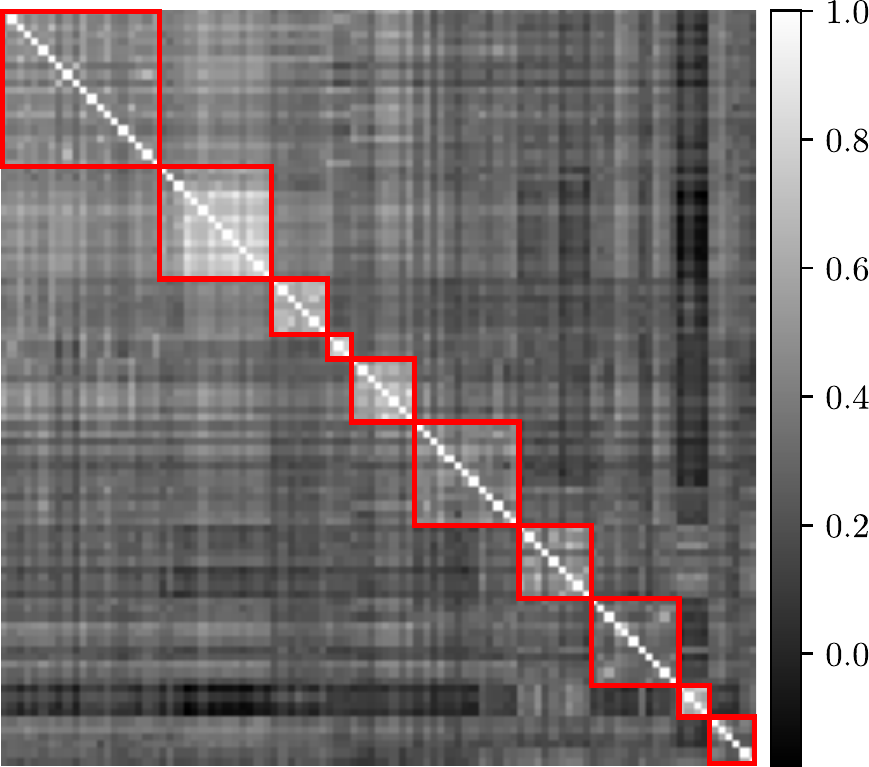};

    \nextgroupplot[
      title={\Large Stock model order selection},  
      legend cell align={left},
      legend style={at={(0.97,0.03)}, anchor=south east},
      xmin=62, xmax=790,
      ylabel={\Large Predicted order \(\displaystyle k\)},
      ymin=0, ymax=11
    ]
    \addplot [semithick, asublue, mark=*, mark size=3, mark options={solid}, forget plot]
    table {%
      95 3.24
      168 5.06
      241 5.86
      314 7.18
      387 8.24
      461 8.56
      534 8.84
      607 9.38
      680 9.72
      753 10
      753 10
    };
    
    
    \addplot [thick, red]
    table {%
      62.1 10
      785.9 10
    };
    \addlegendentry{True order}

    \nextgroupplot[
      title={\Large Stock partition recovery},
      legend cell align={left},
      legend style={draw=white!80.0!black},
      xmin=62, xmax=790,
      ylabel={\Large Error rate},
      ymin=0.00, ymax=0.68
    ]
    \addplot [semithick, asublue, mark=*, mark size=3, mark options={solid}]
    table {%
      95 0.625052631578947
      168 0.490315789473684
      241 0.418947368421053
      314 0.323157894736842
      387 0.246736842105263
      461 0.223157894736842
      534 0.208
      607 0.173894736842105
      680 0.140421052631579
      753 0.0208421052631579
      753 0.0206315789473684
    };
    \addlegendentry{Order estimation}
    \addplot [semithick, asured, mark=square, mark size=3, mark options={solid,fill opacity=0}]
    table {%
      95 0.385052631578947
      168 0.285684210526316
      241 0.24021052631579
      314 0.208842105263158
      387 0.190736842105263
      461 0.170315789473684
      534 0.145052631578947
      607 0.136421052631579
      680 0.107578947368421
      753 0.0275789473684211
      753 0.0246315789473684
    };
    \addlegendentry{Fixed order}

  \end{groupplot}
\end{tikzpicture}